\documentclass[12pt]{imsart}
\usepackage{amsthm,amsmath}
\usepackage[colorlinks,citecolor=blue,urlcolor=blue]{hyperref}

\usepackage{amsmath,amsfonts,amssymb,amsthm}
\usepackage[utf8]{inputenc}
\usepackage{enumitem}

\RequirePackage{amsthm,amsmath,amsfonts,amssymb}
\RequirePackage[numbers]{natbib}
\RequirePackage[colorlinks,citecolor=blue,urlcolor=blue]{hyperref}
\RequirePackage{graphicx}
\usepackage{lineno,hyperref}

\startlocaldefs
\modulolinenumbers[5]

\usepackage[utf8]{inputenc}
\usepackage[english]{babel}

\usepackage[margin=1in]{geometry}
\usepackage{bm}
\usepackage{graphicx}
\usepackage{amssymb,amsmath,amsthm,mathrsfs}
\usepackage{epstopdf}
\usepackage{algorithm}
\usepackage{amsfonts,delarray}
\usepackage{algorithm,algcompatible,amsmath}
\usepackage{rotating,color}
\usepackage{subfigure}
\usepackage{multirow}
\usepackage{titlesec,textcase}
\usepackage{longtable}
\usepackage{bbm}
\usepackage{rotating}

\newtheorem{Theorem}{Theorem}[section]
\newtheorem{Lemma}[Theorem]{Lemma}

\newtheorem{Proposition}[Theorem]{Proposition}

\newtheorem{Definition}[Theorem]{Definition}

\theoremstyle{definition}

\RequirePackage[normalem]{ulem} %DIF PREAMBLE
\definecolor{rp}{RGB}{83,54,106}

\def\boxit#1{\vbox{\hrule\hbox{\vrule\kern6pt\vbox{\kern6pt#1\kern6pt}\kern6pt\vrule}\hrule}}

\endlocaldefs

\allowdisplaybreaks

\begin{document}
\begin{frontmatter}
\title{Empirical likelihood test for  community structure in networks}

\runtitle{EL test for community structure}
%\thankstext{T1}{A sample additional note to the title.}
\runauthor{ }
\begin{aug}

%\author[C]{\fnms{Xiaofeng} \snm{Zhao}\ead[label=e3]{zxfstats@ncwu.edu.cn}},
%\author[B]{\fnms{Wei} \snm{Zhao}\ead[label=e2]{zhaowei2010@outlook.com}}\and
\author[A]{\fnms{Mingao} \snm{Yuan}\ead[label=e1]{mingao.yuan@ndsu.edu}},
\author[B]{\fnms{Sharmin} \snm{Hossain}\ead[label=e2]{sharmin.hossain@ndsu.edu}},
\author[C]{\fnms{Zuofeng} \snm{Shang}\ead[label=e3]{zshang@njit.edu}}

%%%%%%%%%%%%%%%%%%%%%%%%%%%%%%%%%%%%%%%%%%%%%%
%% Addresses                                %%
%%%%%%%%%%%%%%%%%%%%%%%%%%%%%%%%%%%%%%%%%%%%%%
%\address[C]{School of Mathematics and Statistics,
%North China University of Water Resources and Electric Power,
%\printead{e3}}

%\address[B]{Department of Statistics,
%North Dakota State University,
%\printead{e2}}

\address[A]{Department of Statistics,
North Dakota State University,
\printead{e1}}

\address[B]{Department of Statistics,
North Dakota State University,
\printead{e2}}

\address[C]{Department of Mathematical Sciences,
New Jersey Institute of Technology,
\printead{e3}}

\end{aug}

\begin{abstract}
Network data, characterized by interconnected nodes and edges, is pervasive
in various domains and has gained significant popularity in recent years. In network data analysis, testing the presence of community structure in a network is one of the important research tasks. Existing tests are mainly developed for unweighted networks. In this paper, we study the problem of testing the existence of community structure in general (either weighted or unweighted) networks. We propose two new tests: the Weighted Signed-Triangle (WST) test and the empirical likelihood (EL) test. Both tests can be applied to weighted or unweighted networks and outperform existing tests for small networks. The EL test may outperform the WST test for small networks. 
\end{abstract}



\end{frontmatter}

\section{Introduction}
\label{S:1}
A network is a powerful mathematical model used to represent and analyze complex
real-life problems. It consists of a set of objects, referred to as nodes, which are
interconnected through edges. These edges, serving as the links between the
nodes, define the relationships and interactions within the network. Mathematically expressing a network can be achieved through the use of an adjacency
matrix. This matrix serves as a representation, where the rows and columns
correspond to the nodes in the network. By assigning numerical values within
the matrix, we can symbolize the presence of edges between the nodes. Depending on the nature of the edges, a network can be classified as either directed or
undirected. In a directed network, the edges have a specific direction, pointing
from one node to another. This implies that information or influence flows in a
particular direction within the network. On the other hand, an undirected network has bidirectional edges, allowing for connections and interactions to occur
in both directions. Whether directed or undirected, networks provide a flexible
framework to model various systems and phenomena, enabling the examination
of their structures, dynamics, and behaviors. For instance, network is used model social networks, where nodes represent individuals or
entities and the edges typically represent friendships, collaborations, interactions, or any other form of social tie between individuals \cite{N01,ZLZ11,CPV07,N03}.
Network models are also used to understand and represent the complex connectivity patterns and interactions within the brain \cite{SIMS21,NBB17}. Nodes in brain networks
typically represent brain regions or neurons, while edges represent connections
or interactions between them. 

A typical feature of many network data
sets is the presence of community structure \cite{F10,A17,AS17,ACB13,LR15,MS16}.
Community structure refers to the existence of distinct clusters or groups of
nodes within a network, wherein nodes belonging to the same group exhibit a
higher density of connections among themselves compared to nodes outside their
respective groups. These groups represent cohesive subsets of nodes that display
stronger internal interactions and weaker connections with nodes outside their
community. The identification and analysis of community structure in a network
provide valuable insights into the organization, functionality, and dynamics of
complex systems \cite{SIMS21,NBB17,NM21,CY06,AJC15,FYZS18}. By understanding the patterns and relationships within and
between communities, we can gain a deeper understanding of the network’s
behavior and potentially uncover meaningful substructures or functional units
within the larger network.

When examining community structure in networks, hypothesis testing is
commonly employed to assess the significance of the observed network structure \cite{GL17a,JKL18,JKL21,BS16,YYS22,YS22,VA15,AV14,BM17,L16}.
The null
hypothesis assumes the absence of any community structure, while the alternative hypothesis assumes community structure exists. Rejection of the null hypothesis implies the statistical significance of the detected community structure in the network. There are several tests available for this problem. For example, \cite{BS16} proposed the largest eigen-value test and \cite{VA15,AV14,GL17a,JKL18,JKL21} proposed the subgraph-count tests.

Empirical likelihood is a statistical method that provides an alternative approach to traditional likelihood-based inference \cite{O90,O01}. It is a non-parametric statistical method used
to estimate probabilities or construct confidence intervals based on observed
data without making specific assumptions about the underlying probability distribution. By
directly utilizing the empirical distribution of the data, empirical likelihood provides valid statistical inference
even when the data are not normally distributed or when the sample size is
small. Since empirical likelihood method can be used for constructing confidence intervals, hypothesis testing, and model selection, it has applications in various fields, including biostatistics, econometrics,
and machine learning \cite{NS04,PA13,DLL19,RRSPD23,ZY12}.

 In practice, many real-world networks are weighted by interaction frequency, volume or similarity and so on \cite{A14,AJC15, XJL17, ALS18, LWC15,TB11}. Existing  tests for community structure mainly focus on binary (unweighted) networks, where an edge between vertices either exists or not. It is not immediately clear whether  the existing tests can be directly applied to weighted networks or not. For instance, our simulation study shows that the Signed-Triangle (ST) test \cite{JKL21} has little power to detect community structure in small weighted networks (see Section \ref{simu}). One way to solve this issue is to model the edge weights directly. Recently, \cite{YYS22} propose a test by assuming the weights follow some single-parameter distribution. In this paper, we adopt a general degree-corrected stochastic block model to model community structure in weighted networks. This model does not assume any distributional assumption for the weights. We propose two new tests under this model: the Weighted Signed-Triangle (WST) test and the  Empirical Likelihood (EL) test. The EL test and WST test have two advantages over the ST test: (a) the EL test and the WST test can be applied to general (weighted or unweighted) networks; (b) the EL test and the WST test have higher power than the ST test even for small unweighted networks (see Section \ref{simu}). The EL test may outperform the WST test for small networks.

The rest of the paper is organized as follows. Section \ref{mresult} presents the main results. Section \ref{simu} presents simulation study and real data application. The proof is deferred to Section \ref{proof}.

\medskip

\noindent
{\bf Notation:} We adopt the  Bachmann–Landau notation throughout this paper. Let $a_n$  and $b_n$ be two positive sequences. Denote $a_n=\Theta(b_n)$ if $c_1b_n\leq a_n\leq c_2 b_n$ for some positive constants $c_1,c_2$. Denote  $a_n=\omega(b_n)$ if $\lim_{n\rightarrow\infty}\frac{a_n}{b_n}=\infty$. Denote $a_n=O(b_n)$ if $a_n\leq cb_n$ for some positive constants $c$. Denote $a_n=o(b_n)$ if $\lim_{n\rightarrow\infty}\frac{a_n}{b_n}=0$. Let $\mathcal{N}(0,1)$ be the standard normal distribution and $X_n$ be a sequence of random variables. Then $X_n\Rightarrow\mathcal{N}(0,1)$ means $X_n$ converges in distribution to the standard normal distribution as $n$ goes to infinity. 
For positive integer $n$, denote $[n]=\{1,2,\dots,n\}$. Given a finite set $E$, $|E|$ represents the number of elements in $E$. For positive integers $i,j,k$, $i\neq j\neq k$ means $i\neq j$, $j\neq k$, $i\neq k$.

\section{Main results} \label{mresult}

An unweighted graph or network 
 $\mathcal{G}$ consists of a pair $(\mathcal{V},\mathcal{E})$, where $\mathcal{V}=[n]:=\{1,2,\dots,n\}$ denotes the set of vertices and $\mathcal{E}$ denotes the set of edges. 
 For $i<j$, denote $A_{ij}=1$ if $\{i,j\}\in\mathcal{E}$ is an edge and $A_{ij}=0$ otherwise. Suppose  $A_{ii}=0$, that is,  self loops are not allowed. Then the symmetric matrix $A=(A_{ij})\in\{0,1\}^{\otimes n^2}$ is called the adjacency matrix of graph $\mathcal{G}$.  Graph $\mathcal{G}$ is weighted if a weight (a number) is assigned to each edge.
A graph is said to be random if the elements of the adjacency matrix are random.

\begin{Definition}\label{def}
    Let $k_0\ (k_0\geq2)$ be positive integers, $\theta,\tau$ be constants and $Z=(Z_1,\dots,Z_n) $ be independent  uniform random variables on $\{1,2,\dots,k_0\}$.  Given a sequence of positive numbers $W=(w_1,\dots,w_n)$, 
     the degree-corrected weighted stochastic block model $\mathcal{G}(p_n,W, \theta,\tau)$ is defined as follows:  $A_{ij}=\xi_{ij}B_{ij}$, where $\xi_{ij}$ are independent Bernoulli random variables with success probability $p_n$, $B_{ij}$ are conditionally independent given $Z$,
\[\mathbb{E}\big[B_{ij}|Z\big]=w_iw_j\theta_{ij},\hskip 1cm \theta_{ij}=\theta+\tau \big(k_0I[Z_i=Z_j]-1\big),\]
and the tenth moments of $B_{ij}$ are uniformly bounded.
\end{Definition}

In $\mathcal{G}(p_n,W, \theta,\tau)$, the random vector $Z$ randomly assigns community label to each node. Two nodes with the same label $t\in\{1,2,\dots,k_0\}$ belong to the same community labelled as $t$. If two nodes $i,j$ have the same label, that is, $Z_i=Z_j$, then $\mathbb{E}[A_{ij}|Z]=p_nw_iw_j(\theta+(k_0-1)\tau)$. If $Z_i\neq Z_j$, then $\mathbb{E}[A_{ij}|Z]=p_nw_iw_j(\theta-\tau)$. The conditional mean for two nodes within the same community are different from the mean across communities if $\tau\neq 0$. Hence the parameter $\tau$ models the presence of community structure. Note that $\mathbb{E}[A_{ij}]=p_nw_iw_j\theta$ and the expected weighted degree of node $i$ is $\theta p_n w_i\sum_{j\neq i}w_j$. Hence the vector $W$ models the heterogeneity of the means of edges and degrees. The parameter $p_n$ models the existence of edges. If $p_n=0$, all the nodes are isolated. The random variables $B_{ij}$ represent random weights assigned to edges. The condition that the tenth moments of $B_{ij}$ are uniformly bounded is not restrictive. It is required for theoretical analysis in this paper. 

The model $\mathcal{G}(p_n,W, \theta,\tau)$ is pretty general. It includes many existing weighted and unweighted stochastic block models .
When $p_n=1$ and $B_{ij}$ follow the Bernoulli distributions, $\mathcal{G}(p_n,W, \theta,\tau)$ is the unweighted degree-corrected stochastic model \cite{ZLZ12,GMZZ18}. If $B_{ij}\in[0,1]$ and $w_i=1$ $(1\leq i\leq n)$, then $\mathcal{G}(p_n,W, \theta,\tau)$ is the model proposed in \cite{ALS18}. In \cite{FYZS18,NM21}, the weights $B_{ij}$ are assumed to follow the normal distributions and the Gamma distributions respectively.

Given a graph $A\sim \mathcal{G}(p_n,W, \theta,\tau)$, we are interested in testing the following hypotheses:
\begin{equation}\label{hypothesis}
H_0: \ \tau=0,\hskip 2cm H_1: \ \tau\neq0.
\end{equation}
Under $H_0$, the graph  $A$ does not contain  community structure. Under $H_1$, there is community structure in $A$.

The hypothesis problem (\ref{hypothesis}) has been widely studied in the unweighted case and some special weighted case. In the unweighted graph case, \cite{BS16} proposed a test by using the largest eigen-value of a function of the adjacency matrix. 
\cite{GL17a, JKL18} constructed a powerful test by using subgraph counts. Recently, \cite{JKL21} proposed signed polygon tests that can achieve the optimal phase diagram. In the weighted graph case, 
\cite{YYS22} proposed a test under the assumption that the weights follow some single-parameter distributions. \cite{YS22} derive
the sharp information-theoretic limit for the existence of consistent test when the weights follow distributions in the exponential family.

In this paper, we study the hypothesis problem (\ref{hypothesis}) under the model $\mathcal{G}(p_n,W, \theta,\tau)$. Firstly we will extend the Signed Triangle test in \cite{JKL21} to weighted case. Then we propose a test by using  empirical likelihood. As a counterpart of the traditional likelihood, empirical likelihood has been proven to be a powerful and robust statistical inference method \cite{O90,O01,NS04,PA13,DLL19,RRSPD23}. It usually provides valid statistical inference
when the sample size is
small.

Let $d=\sum_{i\neq j}A_{ij}$, $d_i=\sum_{j\neq i}A_{ij}$, and $b_i=\frac{d_i}{\sqrt{d}}$. Denote 
\[T_{ijk}=(A_{ij}-b_ib_j)(A_{jk}-b_jb_k)(A_{ki}-b_kb_i), \ \ i<j<k.\]
If the graph $A$ is unweighted, $T_{ijk}$
 are called signed triangles in \cite{JKL21}. They are used to construct the Signed-Triangle (ST) test. Based on our simulation study (see Section \ref{simu}), the ST test in \cite{JKL21} has little power to detect community structure in small weighted networks. This motivate us to  propose new tests. 
 
 Firstly, we modify the Signed Triangle test in \cite{JKL21} to construct a new powerful test, denoted as the Weighted Signed Triangle (WST) test. To this end, let 
\begin{equation}
 \mathcal{T}_n=\frac{\sum_{i<j<k}T_{ijk}}{\sqrt{\sum_{i<j<k}T_{ijk}^2}}.
\end{equation}
The test statistic $\mathcal{T}_n$ is a counterpart of the signed triangle test statistic $T_n$ in \cite{JKL21}. The only difference lies in the denominator.  Actually, the denominator of $\mathcal{T}_n$ is asymptotically equivalent to the denominator of the ST test statistic in the unweighted network case. However, this small difference makes the WST test much more powerful than the ST test for small weighted or unweighted networks (see Section \ref{simu}). 

\begin{Theorem}\label{nsizeh0}
    In $\mathcal{G}(p_n,W, \theta,\tau)$, suppose $np_n=\omega(1)$ and $c_1\leq \min_iw_i\leq \max_iw_i\leq c_2$ for two positive constants $c_1,c_2$. Then $ \mathcal{T}_n$ converges in distribution under $H_0$ to the standard normal distribution. 
\end{Theorem} 

Based on Theorem \ref{nsizeh0}, the WST test rejects $H_0$ if $|\mathcal{T}_n|\geq Z_{\frac{\alpha}{2}}$, where $Z_{\frac{\alpha}{2}}$ is the $100(1-\frac{\alpha}{2})\%$ quantile of the standard normal distribution and $\alpha$ is the given nominal type I error.

\begin{Theorem}\label{nsizeh1}
    In $\mathcal{G}(p_n,W, \theta,\tau)$, suppose $np_n=\omega(1)$ and $c_1\leq \min_iw_i\leq \max_iw_i\leq c_2$ for two positive constants $c_1,c_2$. Then the power of Weighted Signed Triangle test goes to one if
    $\sqrt{(np_n)^3}(k_0-1)\tau^3=\omega(1)$.
\end{Theorem}

Under the assumptions of Theorem \ref{nsizeh1}, the WST test is consistent if $\sqrt{(np_n)^3}(k_0-1)\tau^3=\omega(1)$. The parameters $k_0,\tau,p_n$ jointly control the power of the test. The power increases as one of the parameters increases.

 Next we construct a new test by using empirical likelihood.
Define the empirical likelihood ratio as 
\begin{equation}\label{el}
\mathcal{R}_n=\max\left\{\prod_{i<j<k}\frac{n(n-1)(n-2)\pi_{ijk}}{6}\Bigg|\sum_{ i<j<k}\pi_{ijk}T_{ijk}=0,\sum_{i<j<k}\pi_{ijk}=1,\pi_{ijk}\geq0\right\}.
\end{equation}
By a Lagrange multiplier argument, the optimal weights $\pi_{ijk}$ of (\ref{el}) are given by
\begin{equation}\label{eleq1}
   \pi_{ijk}=\frac{6}{n(n-1)(n-2)} \frac{1}{1+\lambda T_{ijk}},
\end{equation}
where $\lambda$ is the solution to  
\begin{equation}\label{eleq2}
    \frac{6}{n(n-1)(n-2)}\sum_{ i<j<k}\frac{T_{ijk}}{1+\lambda T_{ijk}}=0.
\end{equation}

The equation (\ref{eleq2}) is non-linear and there is no closed-form solution for $\lambda$ in general. In practice, we can find a numeric approximation of the solution. The quantity $-2\log \mathcal{R}_n$ can be used as a test statistic for (\ref{hypothesis}). It is a counterpart of the log-likelihood ratio in the parametric case. 

\begin{Theorem}\label{sizeho}
   In $\mathcal{G}(p_n, W, \theta,\tau)$, suppose $np_n=\omega(1)$ and $c_1\leq \min_iw_i\leq \max_iw_i\leq c_2$ for two positive constants $c_1,c_2$. Under $H_0$, 
    $-2\log \mathcal{R}_n$ converges in distribution to $\chi^2_1$ as $n\rightarrow\infty$, where $\chi^2_1$ is the chi-square distribution with degree of freedom one.
\end{Theorem}

The proof of Theorem \ref{sizeho} follows the route of the classic empirical likelihood. However, the proof is not trivial due to the facts that $T_{ijk}\ (1\leq i<j<k)$ are not independent, not bounded and the expectations of $T_{ijk}\ (1\leq i<j<k)$ are not zero. Substantial works are needed to overcome these issues.

Based on Theorem \ref{sizeho}, the EL test rejects the null hypothesis $H_0$ if $-2\log \mathcal{R}_n\geq \chi_1^2(1-\alpha)$, where $\chi_1^2(1-\alpha)$ is the $100(1-\alpha)\%$ quantile of the chi-square distribution $\chi^2_1$ and $\alpha$ is the given nominal type I error.

\begin{Theorem}\label{sizeh1}
    In $\mathcal{G}(p_n,W, \theta,\tau)$, suppose $np_n=\omega(1)$ and $c_1\leq \min_iw_i\leq \max_iw_i\leq c_2$ for two positive constants $c_1,c_2$.  Then the power of the empirical likelihood test goes to one if
    $\sqrt{(np_n)^3}(k_0-1)\tau^3=\omega(1)$.
\end{Theorem}

Under the assumptions of Theorem \ref{sizeh1}, the EL test is consistent if $\sqrt{(np_n)^3}(k_0-1)\tau^3=\omega(1)$. The parameters $k_0,\tau,p_n$ jointly control the power of the test. The power increases as one of the parameters increases.

Based on Theorem \ref{nsizeh1} and Theorem \ref{sizeh1}, the EL test and the WST test should have similar performance for large networks. However, the EL test may outperform the WST test for small networks, as validated in Section \ref{simu}. Both the EL test and the WST test outperform the ST test for small networks.

\section{Simulation and real data application}\label{simu}

In this section, we use simulation to evaluate the performance of the proposed EL test and WST test and apply them to several real-world networks.

\subsection{Simulation}

The type I error is set to be 0.05 in the simulation study. We repeat the simulation 500 times to calculate empirical type I errors and powers. Let $X_i\ (1\leq i\leq n)$ be an i.i.d. sample from the uniform distribution on the interval $[0,1]$. We consider two cases of $w_i$: (a) $w_i=\sqrt{3}/\sqrt{1+X_i^2}$ and (b) $w_i=0.8+X_i$. The other parameters are: $k_0=2$, $\theta\in\{0.10, 0.15, 0.20\}$, $\tau\in\{0, 0.02, 0.04, 0.06\}$, $n\in\{30, 40, 50\}$.

In the first simulation, we consider unweighted networks.  Since the edge weights follow Bernoulli distributions, we set $p_n=1$. The results for case (a) are summarized in Table \ref{simu1} and results for  case (b) are summarized in Table \ref{simu2}. Under the null hypothesis ($\tau=0$), the empirical type I errors of the EL test and the WST are close to 0.05, while the type I errors of the ST test are almost zero. This shows the asymptotic distribution of the ST test is not a good approximation of the distribution of the ST test statistic for small networks.  As $\tau$ increases from zero to 0.06, the powers of tests increase and the maximum powers of the three tests are close to one. This shows the consistency of the tests. For fixed $n$, $\theta$ and $\tau$, the powers of the EL test are larger than or equal to the powers of the WST test. This implies that the EL test may outperform the WST test for small networks. Both the EL test and WST test have higher powers than the ST test.

\begin{table}[H]
\begin{center}
\begin{tabular}{ | p {2 cm} | p {2 cm} | p {2 cm} |p {2 cm} | p {2 cm} | }
   \hline
  \multicolumn{5} { | c | }{$n=30$}\\
    \hline
  & $\tau=0$  & $\tau=0.02$   & $\tau=0.04$   & $\tau=0.06$     \\ 
\hline
  \multicolumn{5} { | c | }{EL Test}\\
  \hline
 $\theta=0.1$ & 0.058  & 0.060& 0.216& 0.686     \\ 
  \hline
 $\theta=0.15$ & 0.048  &   0.054& 0.086 &0.430       \\ 
 \hline
 $\theta=0.2$ & 0.058  & 0.046 &0.086& 0.298     \\ 
 \hline
   \multicolumn{5} { | c | }{WST Test}\\
  \hline
 $\theta=0.1$ &  0.054& 0.060 &0.210& 0.678        \\ 
  \hline
 $\theta=0.15$ & 0.050& 0.054& 0.084 &0.430        \\ 
 \hline
 $\theta=0.2$ & 0.070 &0.046 &0.086& 0.298   \\ 
  \hline
  \multicolumn{5} { | c | }{ST Test}\\
  \hline
 $\theta=0.1$ &  0.00 &   0.008 &0.034 &0.376        \\ 
  \hline
 $\theta=0.15$ & 0.00  &   0.002& 0.000 &0.066        \\ 
 \hline
 $\theta=0.2$ & 0.00  &  0.000 &0.000 &0.004   \\ 
 \hline
   \hline
  \multicolumn{5} { | c | }{$n=40$}\\
\hline
  \multicolumn{5} { | c | }{EL Test}\\
  \hline
 $\theta=0.1$ & 0.044  &  0.064& 0.248& 0.878       \\ 
  \hline
 $\theta=0.15$ & 0.054  &    0.076& 0.158& 0.658     \\ 
 \hline
 $\theta=0.2$ & 0.050  & 0.058& 0.094& 0.554         \\ 
  \hline
  \multicolumn{5} { | c | }{WST Test}\\
  \hline
 $\theta=0.1$ & 0.050  &  0.064& 0.246& 0.878    \\ 
  \hline
 $\theta=0.15$ & 0.002  &     0.076& 0.158 &0.656      \\ 
 \hline
 $\theta=0.2$ & 0.000  &   .058& 0.096& 0.554   \\ 
 \hline
 \hline
  \multicolumn{5} { | c | }{ST Test}\\
  \hline
 $\theta=0.1$ &  0  &  0& 0.062& 0.700     \\ 
  \hline
 $\theta=0.15$ & 0  &  0 &0.004& 0.170       \\ 
 \hline
 $\theta=0.2$ & 0  &  0 &0.000& 0.022      \\ 
 \hline
   \hline
  \multicolumn{5} { | c | }{$n=50$}\\
\hline
  \multicolumn{5} { | c | }{EL test}\\
  \hline
 $\theta=0.1$ & 0.054 &0.070 &0.348& 0.982    \\ 
  \hline
 $\theta=0.15$ & 0.056  & 0.052& 0.240& 0.864 \\ 
 \hline
 $\theta=0.2$ & 0.048 & 0.064& 0.142 &0.768 \\ 
 \hline
  \multicolumn{5} { | c | }{WST test}\\
  \hline
 $\theta=0.1$ & 0.042  & 0.070 &0.344 &0.982    \\ 
  \hline
 $\theta=0.15$ &  0.054 &  0.056 &0.240& 0.864   \\ 
 \hline
 $\theta=0.2$ &   0.058 &  0.064 &0.142& 0.768
  \\ 
 \hline
  \multicolumn{5} { | c | }{ST test}\\
  \hline
 $\theta=0.1$ & 0  &  0 &0.112& 0.944    \\ 
  \hline
 $\theta=0.15$ &  0  &  0 &0.008 &0.444  \\ 
 \hline
 $\theta=0.2$ &   0  &  0 &0.000 &0.064
  \\ 
 \hline
\end{tabular}
\caption{Unweighted networks:  $w_i=\sqrt{3}/\sqrt{1+X_i^2}$.}
\label{simu1}
\end{center}
\end{table}

\begin{table}[H]
\begin{center}
\begin{tabular}{ | p {2 cm} | p {2 cm} | p {2 cm} |p {2 cm} | p {2 cm} | }
   \hline
  \multicolumn{5} { | c | }{$n=30$}\\
\hline
  \multicolumn{5} { | c | }{EL Test}\\
  \hline
 $\theta=0.1$ &   0.056 & 0.068 & 0.122 & 0.482      \\ 
  \hline
 $\theta=0.15$ &  0.058&  0.098&  0.074&  0.258       \\ 
 \hline
 $\theta=0.2$ &   0.056&  0.056&  0.066 & 0.188         \\ 
 \hline
  \multicolumn{5} { | c | }{WST Test}\\
  \hline
 $\theta=0.1$ &  0.056 & 0.060&  0.112 & 0.470        \\ 
  \hline
 $\theta=0.15$ &  0.050 & 0.098&  0.074&  0.258       \\ 
 \hline
 $\theta=0.2$ & 0.058&  0.056&  0.066 & 0.188         \\ 
 \hline
   \multicolumn{5} { | c | }{ST Test}\\
 \hline
  $\theta=0.1$ &   0.010 & 0.006&  0.022 & 0.244       \\ 
  \hline
 $\theta=0.15$ & 0.002&  0.000 & 0.008 & 0.040       \\ 
 \hline
 $\theta=0.2$ & 0.000 & 0.000 & 0.004 & 0.006         \\ 
 \hline
   \hline
  \multicolumn{5} { | c | }{$n=40$}\\
\hline
  \multicolumn{5} { | c | }{EL Test}\\
  \hline
 $\theta=0.1$ &  0.048& 0.082 &0.206 &0.738        \\ 
  \hline
 $\theta=0.15$ & 0.050 &0.054& 0.140& 0.464       \\ 
 \hline
 $\theta=0.2$ &  0.058& 0.044& 0.080 &0.352         \\ 
 \hline
  \multicolumn{5} { | c | }{WST Test}\\
  \hline
 $\theta=0.1$ &   0.042 &0.080& 0.20& 0.732       \\ 
  \hline
 $\theta=0.15$ &  0.048& 0.054 &0.14& 0.454     \\ 
 \hline
 $\theta=0.2$ &  0.058& 0.044& 0.08 &0.348         \\ 
 \hline
   \multicolumn{5} { | c | }{ST Test}\\
 \hline
  $\theta=0.1$ & 0.01& 0.012& 0.044& 0.516      \\ 
  \hline
 $\theta=0.15$ & 0.00 &0.000 &0.004& 0.088        \\ 
 \hline
 $\theta=0.2$ &  0.00& 0.000& 0.000 &0.004         \\ 
 \hline
   \hline
  \multicolumn{5} { | c | }{$n=50$}\\
\hline
  \multicolumn{5} { | c | }{EL Test}\\
  \hline
 $\theta=0.1$ & 0.050 &0.078& 0.210& 0.802       \\ 
  \hline
 $\theta=0.15$ & 0.044& 0.046& 0.104& 0.498       \\ 
 \hline
 $\theta=0.2$ & 0.054 & 0.050& 0.078& 0.326         \\ 
 \hline
  \multicolumn{5} { | c | }{WST Test}\\
  \hline
 $\theta=0.1$ &  0.054& 0.076 &0.206 &0.796       \\ 
  \hline
 $\theta=0.15$ & 0.044& 0.046& 0.102& 0.492       \\ 
 \hline
 $\theta=0.2$ &  0.052& 0.050& 0.078 &0.326        \\ 
 \hline
   \multicolumn{5} { | c | }{ST Test}\\
 \hline
  $\theta=0.1$ &  0.008& 0.022 &0.082 &0.616       \\ 
  \hline
 $\theta=0.15$ &   0.004& 0.004& 0.008 &0.134         \\ 
 \hline
 $\theta=0.2$ & 0.000& 0.002 &0.000& 0.028         \\ 
 \hline
\end{tabular}
\caption{Unweighted networks:   $w_i=0.8+X_i$.}
\label{simu2}
\end{center}
\end{table}

In the second simulation, we consider weighted networks, where the weights are generated from the Beta distribution with probability density function
\[f(x)=\frac{1}{B(\alpha,\beta)}x^{\alpha-1}(1-x)^{\beta -1},\]
and $B(\alpha,\beta)$ is the Beta function. The mean $\mu$ of the Beta distribution is equal to $\mu=\frac{\alpha}{\alpha+\beta}$. Let $\mu_{ij}=w_iw_j\theta_{ij}$, where $\theta_{ij}$ is given in Definition \ref{def}. The weight of a pair of nodes $i,j$ is generated from the Beta distribution with  $\beta=2$ and $\alpha=\frac{\beta \mu_{ij}}{1-\mu_{ij}}$. The edge existence parameter $p_n=0.5$. The results are presented in Table \ref{simu3} and Table \ref{simu4}. The sizes of the ST test are zero. This indicates the standard normal distribution is not a good approximation of the ST test statistic. Hence it is necessary to propose new tests. The powers of the ST test are almost zero, which implies the ST test has little power to detect community structure in small weighted networks.  The sizes of the EL test and the WST test fluctuate around the nominal level 0.05. The limiting distributions of the EL test statistic and the WST test statistic provide good approximation of the distributions of statistics. The pattern of the powers of the EL test and the WST test is similar to the unweighted case.

\begin{table}[H]
\begin{center}
\begin{tabular}{ | p {2 cm} | p {2 cm} | p {2 cm} |p {2 cm} | p {2 cm} |}
  \hline
  \multicolumn{5} { | c | }{$n=30,\ \ p_n=0.5$}\\
\hline
  \multicolumn{5} { | c | }{EL Test}\\
  \hline
  & $l=0$  & $l=0.02$   & $l=0.04$   & $l=0.06$     \\ 
  \hline
 $d=0.1$ &     0.058& 0.074& 0.148& 0.442  \\ 
  \hline
 $d=0.15$ &    0.054& 0.094& 0.094 &0.216    \\ 
 \hline
 $d=0.2$ & 0.048 & 0.062 & 0.090&  0.148     \\ 
 \hline
  \multicolumn{5} { | c | }{WST Test}\\
  \hline
 $d=0.1$ &    0.056 &0.062 &0.148& 0.420    \\ 
  \hline
 $d=0.15$ &  0.048& 0.084& 0.080 &0.208   \\ 
 \hline
 $d=0.2$ &  0.046& 0.062 &0.092 &0.148   \\ 
   \hline
  \multicolumn{5} { | c | }{ST Test}\\
  \hline
 $d=0.1$ &   0  &  0  &  0 &0.002    
  \\ 
  \hline
 $d=0.15$ &   0  &  0  &  0 &0.000    \\ 
 \hline
 $d=0.2$ &   0 &   0 &   0 &0.000   \\ 
 \hline
 \hline
  \multicolumn{5} { | c | }{$n=40,\ \ p_n=0.5$}\\
\hline
  \multicolumn{5} { | c | }{EL Test}\\
  \hline
 $d=0.1$ & 0.060  & 0.080 & 0.222&  0.684        \\ 
  \hline
 $d=0.15$ & 0.058  &   0.074&  0.126&  0.340      \\ 
 \hline
 $d=0.2$ & 0.056  &   0.088&  0.068&  0.176        \\ 
 \hline
  \hline
  \multicolumn{5} { | c | }{WST Test}\\
  \hline
 $d=0.1$ & 0.052  &   0.070&  0.208&  0.672       \\ 
  \hline
 $d=0.15$ & 0.052  &   0.068&  0.120&  0.336      \\ 
 \hline
 $d=0.2$ & 0.058  &   0.088 & 0.062&  0.166      \\ 
 \hline
\hline
  \multicolumn{5} { | c | }{ST Test}\\
  \hline
 $d=0.1$ &   0  &  0  &  0 &0.000    
  \\ 
  \hline
 $d=0.15$ &   0  &  0  &  0 &0.000    \\ 
 \hline
 $d=0.2$ &   0 &   0 &   0 &0.000   \\ 
 \hline
  \multicolumn{5} { | c | }{$n=50,\ \ p_n=0.5$}\\
\hline
  \multicolumn{5} { | c | }{EL Test}\\
  \hline
 $d=0.1$ &  0.054 &0.100& 0.328& 0.930      \\ 
  \hline
 $d=0.15$ &  0.050& 0.064& 0.122& 0.544       \\ 
 \hline
 $d=0.2$ & 0.06& 0.068& 0.088& 0.288    \\ 
 \hline
  \hline
  \multicolumn{5} { | c | }{WST Test}\\
  \hline
 $d=0.1$ &  0.052 &0.096& 0.316 &0.928      \\ 
  \hline
 $d=0.15$ &  0.050& 0.064 &0.120& 0.538      \\ 
 \hline
 $d=0.2$ & 0.06& 0.066& 0.088& 0.282    \\ 
 \hline
\hline
  \multicolumn{5} { | c | }{ST Test}\\
  \hline
 $d=0.1$ &   0  &  0  &  0 &0.010    
  \\ 
  \hline
 $d=0.15$ &   0  &  0  &  0 &0.000    \\ 
 \hline
 $d=0.2$ &   0 &   0 &   0 &0.000   \\ 
 \hline
\end{tabular}
\caption{Weighted networks: $w_i=0.8+X_i$, $\beta=2$.}
\label{simu3}
\end{center}
\end{table}

\begin{table}[H]
\begin{center}
\begin{tabular}{ | p {2 cm} | p {2 cm} | p {2 cm} |p {2 cm} | p {2 cm} |}
  \hline
  \multicolumn{5} { | c | }{$n=30,\ \ p_n=0.5$}\\
\hline
  \multicolumn{5} { | c | }{EL Test}\\
  \hline
  & $l=0$  & $l=0.02$   & $l=0.04$   & $l=0.06$     \\ 
  \hline
 $d=0.1$  & 0.054& 0.082 &0.244& 0.768      \\ 
  \hline
 $d=0.15$  & 0.056& 0.078& 0.104& 0.378      \\ 
 \hline
 $d=0.2$ & 0.056& 0.078 &0.066& 0.184      \\ 
 \hline
  \multicolumn{5} { | c | }{WST Test}\\
  \hline
 $d=0.1$  & 0.052& 0.076& 0.236& 0.756       \\ 
  \hline
 $d=0.15$  &  0.054& 0.078& 0.102& 0.374       \\ 
 \hline
 $d=0.2$ &  0.058& 0.078 &0.066& 0.184     \\ 
   \hline
  \multicolumn{5} { | c | }{ST Test}\\
  \hline
 $d=0.1$ &   0  &  0  &  0 &0.002    
  \\ 
  \hline
 $d=0.15$ &   0  &  0  &  0 &0.000    \\ 
 \hline
 $d=0.2$ &   0 &   0 &   0 &0.000   \\ 
 \hline
 \hline
  \multicolumn{5} { | c | }{$n=40,\ \ p_n=0.5$}\\
\hline
  \multicolumn{5} { | c | }{EL Test}\\
  \hline
 $d=0.1$  &  0.054 &0.096& 0.372 &0.948      \\ 
  \hline
 $d=0.15$  &  0.048& 0.082& 0.178& 0.622     \\ 
 \hline
 $d=0.2$ & 0.056& 0.082 &0.086& 0.304\       \\ 
 \hline
  \hline
  \multicolumn{5} { | c | }{WST Test}\\
  \hline
 $d=0.1$  &  0.058& 0.096& 0.358 &0.948      \\ 
  \hline
 $d=0.15$  & 0.050& 0.082& 0.174 &0.616       \\ 
 \hline
 $d=0.2$ &  0.056 &0.082 &0.084 &0.304     \\ 
 \hline
\hline
  \multicolumn{5} { | c | }{ST Test}\\
  \hline
 $d=0.1$ &   0  &  0  &  0 &0.028    
  \\ 
  \hline
 $d=0.15$ &   0  &  0  &  0 &0.000    \\ 
 \hline
 $d=0.2$ &   0 &   0 &   0 &0.000   \\ 
 \hline
  \multicolumn{5} { | c | }{$n=50,\ \ p_n=0.5$}\\
\hline
  \multicolumn{5} { | c | }{EL Test}\\
  \hline
 $d=0.1$  & 0.054 &0.094 &0.550 &1.000     \\ 
  \hline
 $d=0.15$  & 0.056& 0.052& 0.204& 0.844      \\ 
 \hline
 $d=0.2$ &  0.056& 0.064 &0.124 &0.490    \\ 
 \hline
 \hline
  \multicolumn{5} { | c | }{WST Test}\\
  \hline
 $d=0.1$ &  0.052 &0.084 &0.542 &1.000   
  \\ 
  \hline
 $d=0.15$ & 0.052& 0.050& 0.200& 0.842     \\ 
 \hline
 $d=0.2$ &  0.056& 0.064 &0.120 &0.488    \\ 
  \hline
  \multicolumn{5} { | c | }{ST Test}\\
  \hline
 $d=0.1$  &  0  &  0  &  0 &0.184         \\ 
  \hline
 $d=0.15$  &  0  &  0 &   0 &0.002        \\ 
 \hline
 $d=0.2$ &   0  &  0  &  0 &0.000    \\ 
 \hline
\end{tabular}
\caption{Weighted networks: $w_i=\sqrt{3}/\sqrt{1+X_i^2}$, $\beta=2$.}
\label{simu4}
\end{center}
\end{table}

\subsection{Real data application}

In this subsection, we apply the proposed EL test and WST test to three unweighted and four weighted real-world networks available in \cite{NDC}. The networks and the number of nodes are listed in Table \ref{real}. We calculate the p-values of the proposed tests and the ST test and report them in Table \ref{real}. The p-values of the EL test are smaller than that of the WST test. For the unweighted networks `road-chesapeake' and `ENZYMES-g143', the EL test and the WST test have p-values less than 0.05, but the ST test has p-values larger than 0.05. The conclusions of the three tests are not consistent. For the weighted networks, the p-values of the ST test are all zeros. Since the ST test is designed for unweighted networks, the result of the ST test may not be reliable in weighted network case. 
The EL test and WST test have p-values greater than 0.05 for networks `eco-stmarks`, `eco-mangwet` and `eco-wm`', with p-value of network `ca-sandi-auths' less than 0.05. The EL test and the WST test produces the same conclusion with type I error 0.05. In summary, the community structures in networks `soc-karate`, `road-chesapeake', `ENZYMES-g143' and `ca-sandi-auths' are significant based on the EL test and the WST test.

\begin{table}[H]
\begin{center}
\begin{tabular}{ | p {3 cm} | p {0.5 cm} | p {2 cm} | p {3 cm} | p {3 cm} |p {3 cm} |}
  \hline
 network & $n$ & weight& $p$-value of EL test  &  $p$-value of WST test  &  $p$-value of ST test  \\
 \hline
   soc-karate & 34 & unweighted &  0.000    & 0.000 & 0.024  \\ 
  \hline
  road-chesapeake & 39 & unweighted &  0.006    &   0.029  & 0.185 \\ 
  \hline
  ENZYMES-g143 & 39 & unweighted &  0.033    & 0.033 &  0.051  \\ 
 \hline
  eco-stmarks& 54 & weighted   & 0.105    & 0.161  & 0.000\\ 
 \hline
 ca-sandi-auths & 86 & weighted   & 0.000    & 0.000  & 0.000\\ 
 \hline
  eco-mangwet& 97 &  weighted   & 0.318    & 0.363  & 0.000\\
  \hline
  eco-wm1& 277 & weighted   & 0.632    & 0.680  & 0.000\\ 
 \hline
\end{tabular}
\caption{p-values of the EL test, WST test and ST test.}\label{real}
\end{center}
\end{table}

\section{Proof of main results}\label{proof}

The proofs of Theorem \ref{nsizeh0} and Theorem \ref{nsizeh1} follow from the proofs of Theorem \ref{sizeho} and Theorem \ref{sizeh1} directly. Hence we only provide detailed proofs of Theorem \ref{sizeho} and Theorem \ref{sizeh1}. To unify the proofs, we shall define a new empirical likelihood given $A\sim \mathcal{G}(p_n,W, \theta,\tau)$.

Suppose $A\sim \mathcal{G}(p_n,W, \theta,\tau)$. Recall that $d=\sum_{i,j}A_{ij}$, $d_i=\sum_{j}A_{ij}$, $b_i=\frac{d_i}{\sqrt{d}}$ and
\[T_{ijk}=(A_{ij}-b_ib_j)(A_{jk}-b_jb_k)(A_{ki}-b_kb_i).\]
 Let $\eta_{ij}=w_iw_jp_n\big(\theta+\tau \big(k_0I[Z_i=Z_j]-1\big)\big)$. Then $\mathbb{E}[A_{ij}|Z]=\eta_{ij}$ and 
\begin{equation}\label{etah}
\eta_{ij}-\beta_i\beta_j=w_iw_jp_n\tau \big(k_0I[Z_i=Z_j]-1\big).
\end{equation} 
Let $\mu=\mathbb{E}[d]=\Theta(n^2p_n)$, $\mu_i=\mathbb{E}[d_i]=\Theta(np_n)$ uniformly for all $i$, and $\beta_i=\frac{\mu_i}{\sqrt{\mu}}$. Then $\mathbb{E}[A_{ij}]=\beta_i\beta_j$.

 Denote 
\begin{eqnarray*}
\widetilde{T}_{ijk}&=&T_{ijk}-\eta_{ijk},
\end{eqnarray*}
where
\begin{eqnarray*}
\eta_{ijk}&=&(A_{ij}-\eta_{ij})(\eta_{jk}-\beta_j\beta_k)(\eta_{ki}-\beta_k\beta_i)+(\eta_{ij}-\beta_i\beta_j)(A_{jk}-\eta_{jk})(\eta_{ki}-\beta_k\beta_i)\\
&&+(\eta_{ij}-\beta_i\beta_j)(\eta_{jk}-\beta_j\beta_k)(A_{ki}-\eta_{ki})+(\eta_{ij}-\beta_i\beta_j)(\eta_{jk}-\beta_j\beta_k)(\eta_{ki}-\beta_k\beta_i)\\
&&-\frac{1}{d}(\eta_{ij}-\beta_i\beta_j)(\eta_{jk}-\beta_j\beta_k)\sum_{l\neq t}(\eta_{kl}-\beta_k\beta_l)(\eta_{it}-\beta_i\beta_t).
\end{eqnarray*}
Define the empirical likelihood as 
\begin{equation}\label{el1}
\widetilde{\mathcal{R}}_n=\max\left\{\prod_{i<j<k}\frac{n(n-1)(n-2)\pi_{ijk}}{6}\Bigg|\sum_{ i<j<k}\pi_{ijk}\widetilde{T}_{ijk}=0,\sum_{i<j<k}\pi_{ijk}=1,\pi_{ijk}\geq0\right\}.
\end{equation}
By a Lagrange multiplier argument, the optimal weights $\pi_{ijk}$ of (\ref{el1}) are given by
\begin{equation}\label{neleq1}
   \pi_{ijk}=\frac{6}{n(n-1)(n-2)} \frac{1}{1+\lambda \widetilde{T}_{ijk}},
\end{equation}
where $\lambda$ is the solution to  
\begin{equation}\label{neleq2}
    \frac{6}{n(n-1)(n-2)}\sum_{ i<j<k}\frac{\widetilde{T}_{ijk}}{1+\lambda \widetilde{T}_{ijk}}=0.
\end{equation}

Note that $\eta_{ij}-\beta_i\beta_j=0$ by (\ref{etah}) under $H_0$. Hence $\eta_{ijk}=0$ and $\widetilde{\mathcal{R}}_n$ is the empirical likelihood defined in (\ref{el}) under $H_0$.

Now we decompose $\widetilde{T}_{ijk}$ as a sum of leading term and reminder terms.
Simple algebra yields
\begin{eqnarray}\nonumber
    A_{ij}-b_ib_j&=&(A_{ij}-\beta_i\beta_j)+\beta_i\beta_j-b_ib_j\\ \label{ppreq1}
    &=&(A_{ij}-\beta_i\beta_j)-(b_i-\beta_i)(b_j-\beta_j)-(b_i-\beta_i)\beta_j-\beta_i(b_j-\beta_j).
\end{eqnarray}
Plugging (\ref{ppreq1}) into $\widetilde{T}_{ijk}$ and straightforward calculation yields
\begin{eqnarray}\nonumber
\widetilde{T}_{ijk}&=&(A_{ij}-\beta_i\beta_j)(A_{jk}-\beta_j\beta_k)(A_{ki}-\beta_k\beta_i)-\eta_{ijk}-R_{ijk},
\end{eqnarray}
where  $R_{ijk}$ is given by
\begin{eqnarray}\label{eq1}
R_{ijk}=R_{1,ijk}+R_{1,jki}+R_{1,kij}+R_{2,ijk}+R_{2,jki}+R_{2,kij}+R_{3,ijk},
\end{eqnarray}
\begin{eqnarray}\nonumber
  R_{1,ijk}&=&(A_{ij}-\beta_i\beta_j)(A_{jk}-\beta_j\beta_k)(b_k-\beta_k)(b_i-\beta_i)+(A_{ij}-\beta_i\beta_j)(A_{jk}-\beta_j\beta_k)(b_k-\beta_k)\beta_i\\ \label{eq2}
      &&+(A_{ij}-\beta_i\beta_j)(A_{jk}-\beta_j\beta_k)(b_i-\beta_i)\beta_k,
\end{eqnarray}
\begin{eqnarray}\nonumber
   R_{2,ijk} &=&(A_{ij}-\beta_i\beta_j)(b_j-\beta_j)(b_k-\beta_k)^2(b_i-\beta_i)+(A_{ij}-\beta_i\beta_j)(b_j-\beta_j)(b_k-\beta_k)^2 \beta_i\\ \nonumber
    &&+(A_{ij}-\beta_i\beta_j)(b_i-\beta_i)(b_j-\beta_j)(b_k-\beta_k)\beta_k\\ \nonumber
    &&+(A_{ij}-\beta_i\beta_j)(b_i-\beta_i)(b_j-\beta_j)(b_k-\beta_k)\beta_k+(A_{ij}-\beta_i\beta_j)(b_j-\beta_j)(b_k-\beta_k)\beta_i\beta_k\\ \nonumber
    &&+(A_{ij}-\beta_i\beta_j)(b_i-\beta_i)(b_j-\beta_j)\beta_k^2\\ \nonumber
&&+(A_{ij}-\beta_i\beta_j)(b_i-\beta_i)(b_k-\beta_k)^2\beta_j+(A_{ij}-\beta_i\beta_j)(b_i-\beta_i)(b_k-\beta_k)^2\beta_i\beta_j\\ \label{eq3}
&&+(A_{ij}-\beta_i\beta_j)(b_i-\beta_i)(b_k-\beta_k)\beta_j\beta_k,
\end{eqnarray}
\begin{eqnarray} \nonumber
   R_{3,ijk} &=&(b_i-\beta_i)^2(b_j-\beta_j)^2(b_k-\beta_k)^2+R_{4,ijk} +R_{4,jki} +R_{4,kij} \\  \label{eq4}
   &&+R_{5,ijk} +R_{5,jki} +R_{5,kij} +R_{6,ijk},
\end{eqnarray}
\begin{eqnarray}\label{eq4}
   R_{4,ijk} &=&(b_i-\beta_i)(b_j-\beta_j)^2(b_k-\beta_k)^2\beta_i+(b_i-\beta_i)^2(b_j-\beta_j)^2(b_k-\beta_k)\beta_k,
\end{eqnarray}
\begin{eqnarray}\nonumber
   R_{5,ijk} &=&(b_i-\beta_i)(b_j-\beta_j)^2(b_k-\beta_k)\beta_i\beta_k+(b_i-\beta_i)^2(b_j-\beta_j)^2\beta_k^2\\ \label{eq5}
   &&+(b_i-\beta_i)(b_j-\beta_j)(b_k-\beta_k)^2\beta_i\beta_j+(b_i-\beta_i)^2(b_j-\beta_j)(b_k-\beta_k)\beta_j\beta_k,
\end{eqnarray}
\begin{eqnarray}\nonumber
   R_{6,ijk} &=&2(b_i-\beta_i)(b_j-\beta_j)(b_k-\beta_k)\beta_i\beta_j\beta_k+ (b_i-\beta_i)^2(b_j-\beta_j)\beta_j\beta_k^2\\ \nonumber
   &&+(b_i-\beta_i)(b_k-\beta_k)^2\beta_i\beta_j^2+(b_i-\beta_i)^2(b_k-\beta_k)\beta_j^2\beta_k\\ \nonumber
   &&+(b_j-\beta_j)^2(b_k-\beta_k)\beta_i^2\beta_k+(b_j-\beta_j)^2(b_i-\beta_i)\beta_i\beta_k^2\\ \label{eq6}
   &&+(b_j-\beta_j)(b_k-\beta_k)^2\beta_i^2\beta_j.
\end{eqnarray}

Next, we present several lemmas before we prove Theorem \ref{sizeho} and Theorem \ref{sizeh1}. The proofs of these lemmas are complex and lengthy. We defer them to subsection \ref{prlem}.

Denote $\sigma_{ij}^2=\mathbb{E}[(A_{ij}-\beta_i\beta_j)^2]$, $\sigma_n^2=\sum_{i<j<k}\sigma_{ij}^2\sigma_{jk}^2\sigma_{ki}^2$ and
\[U_n=\sum_{i<j<k}(A_{ij}-\beta_i\beta_j)(A_{jk}-\beta_j\beta_k)(A_{ki}-\beta_k\beta_i).\]

\begin{Lemma}\label{lemNor}
Under $H_0$ and the assumptions of Theorem \ref{sizeho}, we have
\begin{equation}\label{Normality}
    \frac{U_n}{\sigma_n}\Rightarrow N(0,1).
\end{equation}
\end{Lemma}

\begin{Lemma}\label{lemdmu}
Suppose the assumptions of Theorem \ref{sizeho} hold. Given positive integer $k$ , the following holds uniformly for all $i$:
\begin{eqnarray*}
\mathbb{E}\big[(d_i-\mu_i)^{2k}\big]=O((np_n)^k), \ \ k\leq 5.
\end{eqnarray*}
\end{Lemma}

\begin{Lemma}\label{plem1}
Under the assumptions of Theorem \ref{sizeho}, the following results hold.
    \[\sqrt{d}-\sqrt{\mu}=O_P(1),\ \ \ \ \ d=\mu\left(1+o_P\left(\frac{1}{\sqrt{n^2p_n}}\right)\right).\]
\end{Lemma}

\begin{Lemma}\label{prlem}
Under the assumptions of Theorem \ref{sizeho}, we have
    \begin{eqnarray*}
\sum_{i}(b_i-\beta_i)^m
=O_P\left(\frac{1}{n^{\frac{m}{2}-1}}\right),
\end{eqnarray*}
for positive integer $m\leq 5$. 
\end{Lemma}

\begin{Lemma}\label{rijk}
Under the assumptions of Theorem \ref{sizeho}, we have
    \[\sum_{i\neq j\neq k}R_{ijk}=o_P(np_n\sqrt{np_n}).\]
\end{Lemma}

\begin{Lemma}\label{rijksq}
 Under the assumptions of Theorem \ref{sizeho},
 \begin{eqnarray*} 
\sum_{i\neq j\neq k}R_{ijk}^2=o_P((np_n)^3).
\end{eqnarray*}  
\end{Lemma}

\begin{Lemma}\label{lemvar}
    Under the assumptions of Theorem \ref{sizeho} and $H_0$,
\begin{eqnarray*}
\frac{1}{(np_n)^3}\sum_{i\neq j\neq k}(A_{ij}-\beta_i\beta_j)^2(A_{jk}-\beta_j\beta_k)^2(A_{ki}-\beta_k\beta_i)^2=\frac{1}{(np_n)^3}\sum_{i\neq j\neq k}\sigma_{ij}^2\sigma_{jk}^2\sigma_{ki}^2+o_P(1)
\end{eqnarray*}

\end{Lemma}

\begin{Lemma}\label{lemmax}
Under the assumptions of Theorem \ref{sizeho}, 
$\max_{i<j<k}|\widetilde{T}_{ijk}|=o_P(np_n\sqrt{np_n})$.
\end{Lemma}

\begin{Lemma}\label{lemeta}
Under $H_1$ and the assumptions of Theorem \ref{sizeho}, we have
\[\sum_{i\neq j\neq k}\eta_{ijk}=\Theta(n^3p_n^3\tau^3(k_0-1))+O_P(n^2p_n^2),\]
  \[  \sum_{i\neq j\neq k}(A_{ij}-\beta_i\beta_j)(A_{jk}-\beta_j\beta_k)(A_{ki}-\beta_k\beta_i)=O_P(\sqrt{n^3p_n^3})+\sum_{i\neq j\neq k}\eta_{ijk}.
\]
\end{Lemma}

The proofs of Lemma \ref{lemNor}-Lemma  \ref{lemeta} are pretty complex and lengthy. We defer them to subsection \ref{prooflemma}.

\subsection{Proof of Theorem \ref{sizeho}}

Under $H_0$, $\tau=0$. In this case, $\eta_{ij}-\beta_i\beta_j=0$ by (\ref{etah}). Hence $\widetilde{T}_{ijk}=T_{ijk}$ and  Lemma \ref{lemNor}-Lemma  \ref{lemeta} hold. By (\ref{eleq2}), we get
\begin{eqnarray*}\nonumber
    0&=&\frac{6}{n(n-1)(n-2)p_n^3}\sum_{ i<j<k}\frac{T_{ijk}}{1+\lambda T_{ijk}}\\ \label{emleq1}
    &=&\frac{6}{n(n-1)(n-2)p_n^3}\sum_{ i<j<k}T_{ijk}-\frac{6}{n(n-1)(n-2)p_n^3}\sum_{ i<j<k}\frac{\lambda T_{ijk}^2}{1+\lambda T_{ijk}}. 
\end{eqnarray*}
Taking absolute value on both sides yields
\begin{eqnarray}\label{thm1eq1}
    \Bigg|\frac{6}{n(n-1)(n-2)p_n^3}\sum_{ i<j<k}T_{ijk}\Bigg|\geq\frac{|\lambda|}{1+|\lambda| \max_{i<j<k}|T_{ijk}|}\frac{6}{n(n-1)(n-2)p_n^3}\sum_{ i<j<k}T_{ijk}^2. 
\end{eqnarray}

By Lemma \ref{rijksq} and Lemma \ref{lemvar}, we have
\begin{equation}\label{thm1eq2}
\frac{6}{n(n-1)(n-2)p_n^3}\sum_{ i<j<k}T_{ijk}^2=\Theta(1)+o_P(1).
\end{equation}
Note that
\begin{eqnarray}\label{thm1eq5}
    \sum_{i<j<k}T_{ijk} 
    &=&\sum_{i<j<k}(A_{ij}-\beta_i\beta_j)(A_{jk}-\beta_j\beta_k)(A_{ki}-\beta_k\beta_i)+\sum_{i<j<k}R_{ijk}.
\end{eqnarray}
By Lemma \ref{lemNor}, we have
\begin{equation}\label{thm1eq4}
\frac{\sum_{i<j<k}(A_{ij}-\beta_i\beta_j)(A_{jk}-\beta_j\beta_k)(A_{ki}-\beta_k\beta_i)}{\sigma_n}\Rightarrow N(0,1).
\end{equation}
Combining (\ref{thm1eq5}), (\ref{thm1eq4}) and Lemma \ref{rijk} yields
\begin{equation}\label{thm1eq3}
\frac{6}{n(n-1)(n-2)p_n^3}\sum_{ i<j<k}T_{ijk}=O_P\left(\frac{1}{\sqrt{(np_n)^3}}\right).
\end{equation}

By (\ref{thm1eq1}), (\ref{thm1eq2}) and (\ref{thm1eq3}), we get
\begin{eqnarray}\nonumber
    |\lambda|\left(1-\max_{i<j<k}|T_{ijk}|O_P\left(\frac{1}{\sqrt{(np_n)^3}}\right)\right)=O_P\left(\frac{1}{\sqrt{(np_n)^3}}\right). 
\end{eqnarray}
Then based on Lemma \ref{lemmax}, it follows that 
\begin{eqnarray}\label{lambdasc}
    |\lambda|=O_P\left(\frac{1}{\sqrt{(np_n)^3}}\right). 
\end{eqnarray}

Next we find an asymptotic expression of $\lambda$.
By (\ref{eleq2}), we get
\begin{eqnarray}\nonumber
    0&=&\frac{6}{n(n-1)(n-2)p_n^3}\sum_{ i<j<k}\frac{T_{ijk}}{1+\lambda T_{ijk}}\\  \nonumber
    &=&\frac{6}{n(n-1)(n-2)p_n^3}\sum_{ i<j<k}T_{ijk}-\frac{6\lambda}{n(n-1)(n-2)p_n^3}\sum_{ i<j<k} T_{ijk}^2\\  \label{emleq2}
    &&+\frac{6\lambda^2}{n(n-1)(n-2)p_n^3}\sum_{ i<j<k}\frac{ T_{ijk}^3}{1+\lambda T_{ijk}}.
\end{eqnarray}

By Lemma \ref{lemmax} and (\ref{lambdasc}), we have
\begin{eqnarray}\nonumber
 \left|\frac{6\lambda^2}{n(n-1)(n-2)p_n^3}\sum_{ i<j<k}\frac{ T_{ijk}^3}{1+\lambda T_{ijk}}\right|&\leq& \frac{\lambda^2\max_{i<j<k}|T_{ijk}|}{1-|\lambda|\max_{i<j<k}|T_{ijk}|}\frac{6}{n(n-1)(n-2)p_n^3}\sum_{ i<j<k}T_{ijk}^2\\ \label{emleq3}
 &=&o_P\left(\frac{1}{\sqrt{(np_n)^3}}\right).
\end{eqnarray}

Combining (\ref{emleq2}), (\ref{emleq3}) and Lemma \ref{lemvar} yields
\[\lambda=\frac{1}{\sigma_n^2}\sum_{ i<j<k}T_{ijk}+o_P\left(\frac{1}{\sqrt{(np_n)^3}}\right).\]

Note that
\[|\lambda|^3\sum_{i<j<k}|T_{ijk}|^3\leq|\lambda|^3\max_{i<j<k}|T_{ijk}|\sum_{i<j<k}|T_{ijk}|^2=o_P(1).\]
Then by Lemma \ref{lemNor}, one has
\begin{eqnarray} \nonumber
- 2\log{\mathcal{R}_n} &=&2\sum_{i<j<k}\log{\left( 1+\lambda T_{ijk} \right) }  
\\     \nonumber
&=&2\Big(\ \lambda\sum_{i<j<k} T_{ijk}-\frac{\lambda^2}{2}\sum_{i<j<k}T_{ijk}^2+o_P(1)
 \Big) \\ \nonumber
&=&\frac{1}{\sigma_n^2}\left(\sum_{ i<j<k}T_{ijk}\right)^2+o_P(1)\\ \label{emleq5}
&\Rightarrow&\chi^2_1,
\end{eqnarray}
which completes the proof of Theorem \ref{sizeho}.

\qed

\subsection{Proof of Theorem \ref{sizeh1}}
Under $H_1$, $\tau\neq 0$.
By a similar argument as in the proof of Theorem \ref{sizeho}, it is easy to get
\begin{eqnarray*} \nonumber
- 2\log{\widetilde{\mathcal{R}}_n} &=&2\sum_{i<j<k}\log{\left( 1+\lambda \widetilde{T}_{ijk} \right) }  
\\     \nonumber
&=&2\Big(\ \lambda\sum_{i<j<k} \widetilde{T}_{ijk}-\frac{\lambda^2}{2}\sum_{i<j<k}\widetilde{T}_{ijk}^2+o_P(1)
 \Big) \\ \nonumber
&=&\frac{1}{\sigma_n^2}\left(\sum_{ i<j<k}T_{ijk}-\sum_{ i<j<k}\eta_{ijk}\right)^2+o_P(1)\\
&=&\left(\frac{\sum_{ i<j<k}T_{ijk}}{\sigma_n}\right)^2+\left(\frac{\sum_{ i<j<k}\eta_{ijk}}{\sigma_n}\right)^2\\
&&-2\left(\frac{\sum_{ i<j<k}T_{ijk}}{\sigma_n}\right)\left(\frac{\sum_{ i<j<k}\eta_{ijk}}{\sigma_n}\right)+o_P(1).
\end{eqnarray*}
Note that
$\sigma_n^2=\Theta(n^3p_n^3)$. By Lemma \ref{lemeta}, we have
\[
\sum_{ i<j<k}\eta_{ijk}=\Theta(n^3p_n^3\tau^3(k_0-1))+O_P(n^2p_n^2),
\]
and 
\[\sum_{ i<j<k}T_{ijk}-\sum_{ i<j<k}\eta_{ijk}=O_P(\sqrt{(np_n)^3}).\]
Hence, if $\sqrt{(np_n)^3}\tau^3(k_0-1)=\omega(1)$, the power of the empirical likelihood test goes to one. Then the proof is complete.

\qed

\subsection{Proof of lemmas}\label{prooflemma}

In this subsection, we provide detailed proofs of Lemma \ref{lemNor}-Lemma \ref{lemeta}.

\subsubsection{Proof of Lemma \ref{lemNor}}

We will use the following proposition to prove Lemma \ref{lemNor}.

\begin{Proposition}[(\cite{HH14})]\label{martingale}
 Suppose that for every $n\in\mathbb{N}$ and $k_n\rightarrow\infty$ the random variables $X_{n,1},\dots,X_{n,k_n}$ are a martingale difference sequence relative to an arbitrary filtration $\mathcal{F}_{n,1}\subset\mathcal{F}_{n,2}$ $\subset$ $\dots$ $\subset\mathcal{F}_{n,k_n}$. If (I) $\sum_{i=1}^{k_n}\mathbb{E}(X_{n,i}^2|\mathcal{F}_{n,i-1})\rightarrow 1$ in probability,
 (II) $\sum_{i=1}^{k_n}\mathbb{E}(X_{n,i}^2I[|X_{n,i}|>\epsilon]|\mathcal{F}_{n,i-1})\rightarrow 0$ in probability for every $\epsilon>0$,
\noindent then $\sum_{i=1}^{k_n}X_{n,i}\rightarrow N(0,1)$ in distribution.
\end{Proposition}

\noindent
{\bf Proof of Lemma \ref{lemNor}:}
Let 
\[Y_t=\frac{\sum_{1\leq i<j<k\leq t}(A_{ij}-\beta_i\beta_j)(A_{jk}-\beta_j\beta_k)(A_{ki}-\beta_k\beta_i)}{\sigma_n},\]
for $3\leq t\leq n$, and $Y_2=0$. Then $\{Y_t\}_{t=2}^n$ is a martingale and $Y_n=\frac{U_n}{\sigma_n}$.

Next we use Proposition \ref{martingale} to prove that $Y_n$ converges in distribution to the standard normal distribution. Let $F_t=\{A_{ij},1\leq i<j\leq t\}$ and $X_t=Y_t-Y_{t-1}$ for $t\geq3$. It is easy to verify that
\[X_t=\frac{\sum_{1\leq i<j<k=t}(A_{ij}-\beta_i\beta_j)(A_{jk}-\beta_j\beta_k)(A_{ki}-\beta_k\beta_i)}{\sigma_n}.\]
Since $\mathbb{E}[A_{ki}]=\beta_{k}\beta_{i}$, then
 $\mathbb{E}[X_{t}|F_{t-1}]=0$. Hence, $\{X_t\}_{t=3}^n$ is a martingale difference. 
 
 Now we verify the two conditions in Proposition \ref{martingale}. We check condition $(I)$ first.
It is easy to get $\mathbb{E}\big(Y_{t}|F_{t-1}\big)=Y_{t-1}$. Then by the definition of $X_t$ and the property of conditional expectation, we have
\begin{eqnarray*}
\mathbb{E}\left[\sum_{t=3}^n\mathbb{E}\Big(X^2_{t}|F_{t-1}\Big)\right]&=&\mathbb{E}\left[\sum_{t=3}^n\mathbb{E}\Big((Y^2_{t}-2Y_{t}Y_{t-1}+Y^2_{t-1})|F_{t-1}\Big)\right]\\
&=&\mathbb{E}\left[\sum_{t=3}^n\mathbb{E}\big[(Y^2_{t}-Y^2_{t-1})|F_{t-1}\big]\right]=\mathbb{E}[Y^2_{n}]=1.
\end{eqnarray*}
Next we show that \[\mathbb{E}\Big(\sum_{t=3}^n\mathbb{E}[X_{t}^2|F_{t-1}]\Big)^2=1+o(1).\] 
Given $t\in\{3,4,\dots,n\}$, one has
\begin{eqnarray*}
\mathbb{E}[X_{t}^2|F_{t-1}]
&=&\frac{1}{\sigma_n^2}\sum_{\substack{1\leq i_1<j_1<t\\1\leq i_2<j_2<t}}\mathbb{E}\Big[(A_{i_1j_1}-\beta_{i_1}\beta_{j_1})(A_{j_1t}-\beta_{j_1}\beta_{t})(A_{ti_1}-\beta_{t}\beta_{i_1})\\
&&\times (A_{i_2j_2}-\beta_{i_2}\beta_{j_2})(A_{j_2t}-\beta_{j_2}\beta_{t})(A_{ti_2}-\beta_{t}\beta_{i_2})|F_{t-1}\Big]\\
&=&\frac{1}{\sigma_n^2}\sum_{1\leq i<j<t}(A_{ij}-\beta_i\beta_j)^2\sigma_{jt}^2\sigma_{it}^2.
\end{eqnarray*}
Then we have
\begin{eqnarray*}
\sum_{t=3}^n\mathbb{E}[X_{t}^2|F_{t-1}]
&=&\frac{1}{\sigma_n^2}\sum_{t=3}^n\sum_{1\leq i<j<t}(A_{ij}-\beta_i\beta_j)^2\sigma_{jt}^2\sigma_{it}^2=\frac{1}{\sigma_n^2}\sum_{1\leq i<j<t\leq n}(A_{ij}-\beta_i\beta_j)^2\sigma_{jt}^2\sigma_{it}^2.
\end{eqnarray*}
Note that $A_{ij}$ $(1\leq i<j\leq n)$ are independent under $H_0$. Let $m_{4,ij}=\mathbb{E}\big[(A_{ij}-\beta_i\beta_j)^4\big]$. Since the tenth moments of $B_{ij}$ are uniformly bounded and $A_{ij}=\xi_{ij}B_{ij}$, then $m_{4,ij}=O(p_n)$ uniformly for all $i<j$. Then
\[\mathbb{E}\Big[(A_{ij}-\beta_i\beta_j)^2(A_{kl}-\beta_k\beta_l)^2\Big]=m_{4,ij},\hskip  1cm if\ \{i,j\}=\{k,l\},\]
and 
\[\mathbb{E}\Big[(A_{ij}-\beta_i\beta_j)^2(A_{kl}-\beta_k\beta_l)^2\Big]=\sigma_{ij}^2\sigma_{kl}^2, \hskip 1cm if\ |\{i,j\}\cap\{k,l\}|\leq1.\]
Hence, we have
\begin{eqnarray}\nonumber
\mathbb{E}\Big(\sum_{t=1}^n\mathbb{E}[X_{n,t}^2|F_{t-1}]\Big)^2&=&\frac{1}{\sigma_n^4}\sum_{\substack{1\leq i<j<t\leq n\\1\leq k<l<s\leq n}}\mathbb{E}(A_{ij}-\beta_i\beta_j)^2\sigma_{jt}^2\sigma_{it}^2(A_{kl}-\beta_k\beta_l)^2\sigma_{ks}^2\sigma_{ls}^2\\  \nonumber
&=&\frac{1}{\sigma_n^4}\sum_{\substack{1\leq i<j<t\leq n\\1\leq k<l<s\leq n\\|\{i,j\}\cap\{k,l\}|\leq1}}\mathbb{E}(A_{ij}-\beta_i\beta_j)^2\sigma_{jt}^2\sigma_{it}^2(A_{kl}-\beta_k\beta_l)^2\sigma_{ks}^2\sigma_{ls}^2\\ \nonumber
&&+\frac{1}{\sigma_n^4}\sum_{\substack{1\leq i<j<t\leq n\\1\leq k<l<s\leq n\\ \{i,j\}=\{k,l\}}}\mathbb{E}(A_{ij}-\beta_i\beta_j)^2\sigma_{jt}^2\sigma_{it}^2(A_{kl}-\beta_k\beta_l)^2\sigma_{ks}^2\sigma_{ls}^2\\ \nonumber
&=& \frac{1}{\sigma_n^4}\sum_{\substack{1\leq i<j<t\leq n\\1\leq k<l<s\leq n\\|\{i,j\}\cap\{k,l\}|\leq1}}\sigma_{ij}^2\sigma_{jt}^2\sigma_{it}^2\sigma_{kl}^2\sigma_{ks}^2\sigma_{ls}^2+\frac{1}{\sigma_n^4}\sum_{\substack{1\leq i<j<t\leq n\\1\leq k<l<s\leq n\\ \{i,j\}=\{k,l\}}}m_{4,ij}^2\sigma_{jt}^2\sigma_{it}^2\sigma_{ks}^2\sigma_{ls}^2\\  \nonumber
&=&\frac{1}{\sigma_n^4}\sum_{\substack{1\leq i<j<t\leq n\\1\leq k<l<s\leq n\\|\{i,j\}\cap\{k,l\}|\leq1}}\sigma_{ij}^2\sigma_{jt}^2\sigma_{it}^2\sigma_{kl}^2\sigma_{ks}^2\sigma_{ls}^2+\frac{1}{\sigma_n^4}\sum_{\substack{1\leq i<j<t\leq n\\1\leq k<l<s\leq n\\\{i,j\}=\{k,l\}}}\sigma_{ij}^2\sigma_{jt}^2\sigma_{it}^2\sigma_{kl}^2\sigma_{ks}^2\sigma_{ls}^2\\ \nonumber
&&+\frac{1}{\sigma_n^4}\sum_{\substack{1\leq i<j<t\leq n\\1\leq k<l<s\leq n\\ \{i,j\}=\{k,l\}}}(m_{4,ij}^2\sigma_{jt}^2\sigma_{it}^2\sigma_{ks}^2\sigma_{ls}^2-\sigma_{ij}^2\sigma_{jt}^2\sigma_{it}^2\sigma_{kl}^2\sigma_{ks}^2\sigma_{ls}^2)\\
&=&1+O\Big(\frac{n^4p_n^5}{(np_n)^6}\Big)=1+o(1).
\end{eqnarray}

 Now we check condition $(II)$ in Proposition \ref{martingale}. 
Let $\epsilon$ be a fixed positive constant. By  the Cauchy-Schwarz inequality and Markov's inequality, we have
\begin{eqnarray}\nonumber
&&\mathbb{E}\Big[\sum_{t=3}^n\mathbb{E}\big[X_{t}^2I[|X_{t}|>\epsilon|F_{t-1}\big]\Big]\\    \nonumber
&\leq&\mathbb{E}\Big[\sum_{t=3}^n\sqrt{\mathbb{E}\big[X_{t}^4|F_{t-1}\big]\mathbb{P}[|X_{t}|>\epsilon|F_{t-1}\big]}\Big]\\  \nonumber
&\leq&\mathbb{E}\Bigg[\frac{1}{\epsilon^2\sigma_n^4}\sum_{t=1}^n\mathbb{E}\Big[\Big(\sum_{1\leq i<j<t}(A_{ij}-\beta_i\beta_j)(A_{jk}-\beta_{j}\beta_{k})(A_{ki}-\beta_{k}\beta_{i})\Big)^4\Big|F_{t-1}\Big]\Bigg]\\   \nonumber
&=&\frac{1}{\epsilon^2\sigma_n^4}\sum_{t=3}^n\sum_{\substack{1\leq i_1<j_1<t\\ 1\leq i_2<j_2<t\\ 1\leq i_3<j_3<t\\ 1\leq i_4<j_4<t}}\mathbb{E}\Big[(A_{i_1j_1}-\beta_{i_1}\beta_{j_1})(A_{j_1t}-\beta_{j_1}\beta_{t})(A_{ti_1}-\beta_{t}\beta_{i_1})\\  \nonumber
&&\times(A_{i_2j_2}-\beta_{i_2}\beta_{j_2})(A_{j_2t}-\beta_{j_2}\beta_{t})(A_{ti_2}-\beta_{t}\beta_{i_2})\\
&&\times(A_{i_3j_3}-\beta_{i_3}\beta_{j_3})(A_{j_3t}-\beta_{j_3}\beta_{t})(A_{ti_3}-\beta_{t}\beta_{i_3})\\ \nonumber
&&\times(A_{i_4j_4}-\beta_{i_4}\beta_{j_4})(A_{j_4t}-\beta_{j_4}\beta_{t})(A_{ti_4}-\beta_{t}\beta_{i_4})\Big]\\   \nonumber
&=&\frac{C_1}{\epsilon^2\sigma_n^4}\sum_{t=3}^n\sum_{\substack{1\leq i_1<j_1<t\\ 1\leq i_2<j_2<t}}\mathbb{E}\big[(A_{i_1j_1}-\beta_{i_1}\beta_{j_1})^2(A_{j_1t}-\beta_{j_1}\beta_{t})^2(A_{ti_1}-\beta_{t}\beta_{i_1})^2\\  \nonumber
&&\times(A_{i_2j_2}-\beta_{i_2}\beta_{j_2})^2(A_{j_2t}-\beta_{j_2}\beta_{t})^2(A_{ti_2}-\beta_{t}\beta_{i_2})^2\big]\\   \nonumber
&&+\frac{1}{\epsilon^2\sigma_n^4}\sum_{t=3}^n\sum_{\substack{1\leq i_1<j_1<t}}\mathbb{E}\big[(A_{i_1j_1}-\beta_{i_1}\beta_{j_1})^4(A_{j_1t}-\beta_{j_1}\beta_{t})^4(A_{ti_1}-\beta_{t}\beta_{i_1})^4\big] \\   \nonumber
&=&O\Big(\frac{n^5p_n^6}{\epsilon^2\sigma_n^4}\Big)+O\Big(\frac{n^3p_n^3}{\epsilon^2\sigma_n^2}\Big)=o(1).
\end{eqnarray}
 Then the desired result follows from Proposition \ref{martingale}.

\qed

\subsubsection{Proof of Lemma \ref{lemdmu}}

\noindent
{\bf Proof of Lemma \ref{lemdmu}: } Note that
\begin{eqnarray*}
d_i-\mu_i=\sum_{j\neq i}(A_{ij}-\eta_{ij})+\sum_{j\neq i}(\eta_{ij}-\beta_i\beta_j).
\end{eqnarray*}
It is easy to verify that
\begin{eqnarray}\label{pfeq1}
\mathbb{E}\big[(d_i-\mu_i)^{2k}\big]\leq 2^{2k-1}\left(\mathbb{E}\Big[\Big(\sum_{j\neq i}(A_{ij}-\eta_{ij})\Big)^{2k}\Big]+\mathbb{E}\Big[\Big(\sum_{j\neq i}(\eta_{ij}-\beta_i\beta_j)\Big)^{2k}\right).
\end{eqnarray}

Next, we bound the two terms in (\ref{pfeq1}). Consider the first term first.
\begin{eqnarray*}
\mathbb{E}\Big[\sum_{j\neq i}(A_{ij}-\eta_{ij})\Big]^{2k}
&=&\sum_{j_1,j_2,\dots,j_{2k}\neq i}\mathbb{E}\big[(A_{ij_1}-\eta_{ij_1})\dots(A_{ij_{2k}}-\eta_{ij_{2k}})\big].
\end{eqnarray*}
Recall that $A_{ij}$ are conditionally independent given $Z$ under $H_1$ and independent under $H_0$. Moreover, $\mathbb{E}[A_{ij}|Z]=\eta_{ij}$.
If $j_1\not\in\{j_2,j_3,\dots,j_{2k}\}$, then  
\begin{eqnarray*}
\mathbb{E}\big[(A_{ij_1}-\eta_{ij_1})\dots(A_{ij_{2k}}-\eta_{ij_{2k}})\big]=\mathbb{E}\Big[\mathbb{E}\big[(A_{ij_1}-\eta_{ij_1})|Z\big]\dots(A_{ij_{2k}}-\eta_{ij_{2k}})\Big]=0.
\end{eqnarray*}
Hence $j_1=j_{t_0}$ for some $t_0\in\{2,3,\dots,2k\}$. The same result holds for all $j_t$ with $t\in\{1,2,\dots,2k\}$. Then $|\{j_1,j_2,\dots,j_{2k}\}|\leq k$. There are at most $n^k$ choices of such indices. Since the tenth moments of $B_{ij}$ are uniformly bounded and $A_{ij}=\xi_{ij}B_{ij}$, then $\mathbb{E}[A_{ij}^{10}]=O(p_n)$ uniformly for all $i<j$. Given positive integer $2\leq r\leq10$, it is easy to verify that 
\begin{eqnarray*}
\Big|\mathbb{E}\big[(A_{ij}-\eta_{ij})^r|Z\big]\Big|\leq \sum_{t=0}^r\binom{r}{t}\mathbb{E}\big[|A_{ij}|^{r-t}|Z\big]\eta_{ij}^{t}=O(p_n).
\end{eqnarray*}
Hence we get
\begin{eqnarray}\label{pfeq2}
\mathbb{E}\Big[\Big(\sum_{j\neq i}(A_{ij}-\eta_{ij})\Big)^{2k}\Big]
&=&O((np_n)^k),\ \ k\leq 5,
\end{eqnarray}
uniformly for all $i$.

Now we bound the second term in (\ref{pfeq1}).
Note that
\begin{eqnarray}\label{ezeq1}
\mathbb{E}\big[(\eta_{ij}-\beta_i\beta_j)|Z_j]=w_iw_jp_n\tau \big(k_0\mathbb{E}\big[I[Z_i=Z_j]\big|Z_j]-1\big)=w_iw_jp_n\tau \big(1-1\big)=0,
\end{eqnarray}
and
\begin{eqnarray*}
\mathbb{E}\Big[\Big(\sum_{j\neq i}(\eta_{ij}-\beta_i\beta_j)\Big)^{2k}\Big]=\sum_{j_1,\dots,j_{2k}}\mathbb{E}\Big[(\eta_{ij_1}-\beta_i\beta_{j_1})\dots(\eta_{ij_1}-\beta_i\beta_{j_{2k}})\Big].
\end{eqnarray*}
If $j_1\not\in\{j_2,j_3,\dots,j_{2k}\}$, then 
\begin{eqnarray*}
&&\mathbb{E}\Big[(\eta_{ij_1}-\beta_i\beta_{j_1})\dots(\eta_{ij_1}-\beta_i\beta_{j_{2k}})\Big]\\
&=&\mathbb{E}\Big[\mathbb{E}\Big[(\eta_{ij_1}-\beta_i\beta_{j_1})\big|Z_i,Z_{j_2},\dots,Z_{j_{2k}}\Big]\dots(\eta_{ij_1}-\beta_i\beta_{j_{2k}})\Big]\\
&=&0.
\end{eqnarray*}
Hence, $j_1=j_{t_0}$ for some $t_0\in\{2,3,\dots,2k\}$. Similar result holds for each $j_t$ with $t=1,2,\dots, 2k$. As a result, $|\{j_1,j_2,\dots,j_{2k}\}|\leq k$. There are at most $n^k$ choices for such indices. Then
\begin{eqnarray}\label{pfeq3}
\mathbb{E}\Big[\Big(\sum_{j\neq i}(\eta_{ij}-\beta_i\beta_j)\Big)^{2k}\Big]=O\left(\sum_{j_1,\dots,j_{k}}\mathbb{E}\Big[(\eta_{ij_1}-\beta_i\beta_{j_1})^2\dots(\eta_{ij_1}-\beta_i\beta_{j_{k}})^2\Big]\right)=O((np_n)^k),
\end{eqnarray}
uniformly for all $i$.

Combining (\ref{pfeq1}), (\ref{pfeq2}) and (\ref{pfeq3}) yields
\begin{eqnarray*}
\mathbb{E}\big[(d_i-\mu_i)^{2k}\big]=O((np_n)^k),\ \ k\leq 5.
\end{eqnarray*}
uniformly for all $i$.

\qed

\subsubsection{Proof of Lemma \ref{plem1}}

\noindent
{\bf Proof of Lemma \ref{plem1}: } By the proof of Lemma \ref{lemdmu}, we have
\begin{eqnarray*}
    d-\mu=\sum_{i\neq j}(A_{ij}-\eta_{ij})+\sum_{i\neq j}(\eta_{ij}-\beta_i\beta_j).
\end{eqnarray*}
Since $A_{ij}$ are conditionally independent given $Z$ and $\mathbb{E}[A_{ij}|Z]=\eta_{ij}$, then
\begin{eqnarray*}
\mathbb{E}\Big[\Big(\sum_{i<j}(A_{ij}-\eta_{ij})\Big)^2\Big]=\mathbb{E}\Big[\sum_{i<j}\mathbb{E}\big[(A_{ij}-\eta_{ij})^2|Z\big]\Big]=O(n^2p_n).
\end{eqnarray*}

For distinct indices $i\neq j\neq k$, one has
\begin{eqnarray*}
\mathbb{E}\big[(\eta_{ij}-\beta_i\beta_j)(\eta_{ik}-\beta_i\beta_k)\big]=\mathbb{E}\Big[(\eta_{ij}-\beta_i\beta_j)\mathbb{E}\big[(\eta_{ik}-\beta_i\beta_k)|Z_i,Z_j\big]\Big]=0.
\end{eqnarray*}
Then
\begin{eqnarray*}
\mathbb{E}\Big[\Big(\sum_{i\neq j}(\eta_{ij}-\beta_i\beta_j)\Big)^2\Big]&=&\sum_{i\neq j}\mathbb{E}\big[(\eta_{ij}-\beta_i\beta_j)^2\big]+\sum_{i\neq j\neq k}\mathbb{E}\big[(\eta_{ij}-\beta_i\beta_j)(\eta_{ik}-\beta_i\beta_k)\big]\\
&=&O(n^2p_n).
\end{eqnarray*}
Hence $d-\mu=O_P\left(\sqrt{n^2p_n}\right)$ and
\begin{eqnarray*}
\sqrt{d}-\sqrt{\mu}=\frac{d-\mu}{\sqrt{d}+\sqrt{\mu}}=O_P(1).
\end{eqnarray*}

\qed

\subsubsection{Proof of Lemma \ref{prlem}}

\noindent
{\bf Proof of Lemma \ref{prlem}:} Given positive integer $m \leq 5$, by the binomial expansion, we have
\begin{eqnarray}\nonumber
\sum_{i}(b_i-\beta_i)^m&=&\sum_{i}\left(\frac{d_i-\mu_i}{\sqrt{d}}-\frac{(\sqrt{d}-\sqrt{\mu})\mu_i}{\sqrt{d\mu}}\right)^m\\ \nonumber
&=&\sum_{k=0}^m(-1)^k\binom{m}{k}\sum_{i}\left(\frac{d_i-\mu_i}{\sqrt{d}}\right)^k\left(\frac{(\sqrt{d}-\sqrt{\mu})\mu_i}{\sqrt{d\mu}}\right)^{m-k}\\ \label{r1ijkeq20}
&=&\sum_{k=0}^m(-1)^k\binom{m}{k}\frac{(\sqrt{d}-\sqrt{\mu})^{m-k}}{\sqrt{d}^{m}\sqrt{\mu}^{m-k}}\sum_{i}(d_i-\mu_i)^k\mu_i^{m-k}
\end{eqnarray}
By the Cauchy-Schwarz inequality and Lemma \ref{lemdmu}, one has
\[ \sum_{i}\mathbb{E}\big[|d_i-\mu_i|^k\big]\leq\sum_{i}\sqrt{\mathbb{E}\big[|d_i-\mu_i|^{2k}\big]}=O(n\sqrt{(np_n)^k}).\]
Then 
\[\sum_{i}(d_i-\mu_i)^k=O_P(n\sqrt{(np_n)^k}),\]
and
\begin{eqnarray}\nonumber
\sum_{i}(b_i-\beta_i)^m=
O_P\left(\max_{0\leq k\leq m}\frac{1}{n^{m-\frac{k}{2}-1}}\right)=O_P\left(\frac{1}{n^{\frac{m}{2}-1}}\right).
\end{eqnarray}

\qed

\subsubsection{Proof of Lemma \ref{rijk}}

\noindent
{\bf Proof of Lemma \ref{rijk}:} The proof proceeds by showing $\sum_{i\neq j\neq k}R_{t,ijk}=o_P(np_n\sqrt{np_n})$ for $t=1,2,3,4,5,6$.

{\bf Step 1: } we prove $\sum_{i\neq j\neq k}R_{1,ijk}=o_P(np_n\sqrt{np_n})$.  We shall bound the summation of each term in $R_{1,ijk}$ over $i\neq j\neq k$.  Simple algebra yields
\begin{eqnarray}\nonumber
    &&\sum_{i\neq j\neq k}(A_{ij}-\beta_i\beta_j)(A_{jk}-\beta_j\beta_k)(b_k-\beta_k)(b_i-\beta_i)\\ \nonumber
    &=&\sum_{i\neq j\neq k}(A_{ij}-\beta_i\beta_j)(A_{jk}-\beta_j\beta_k)\left(\frac{d_k-\mu_k}{\sqrt{d}}-\frac{(\sqrt{d}-\sqrt{\mu})\mu_k}{\sqrt{d\mu}}\right)\left(\frac{d_i-\mu_i}{\sqrt{d}}-\frac{(\sqrt{d}-\sqrt{\mu})\mu_i}{\sqrt{d\mu}}\right)\\ \nonumber
    &=&\frac{1}{d}\sum_{i\neq j\neq k}(A_{ij}-\beta_i\beta_j)(A_{jk}-\beta_j\beta_k)(d_k-\mathbb{E}[d_k])(d_i-\mathbb{E}[d_i])\\ \nonumber
    &&-\frac{2(\sqrt{d}-\sqrt{\mu})}{d\sqrt{\mu}}\sum_{i\neq j\neq k}(A_{ij}-\beta_i\beta_j)(A_{jk}-\beta_j\beta_k)\mu_i(d_k-\mathbb{E}[d_k])\\ \label{r1ijkeq1}
    &&+\frac{(\sqrt{d}-\sqrt{\mu})^2}{d\mu}\sum_{i\neq j\neq k}(A_{ij}-\beta_i\beta_j)(A_{jk}-\beta_j\beta_k)\mu_k\mu_i.
\end{eqnarray}
Next we prove each term in (\ref{r1ijkeq1}) is equal to $o_P(np_n\sqrt{np_n})$.

(a). We prove the first term of (\ref{r1ijkeq1}) is equal to $o_P(np_n\sqrt{np_n})$. Recall that $\eta_{ij}=w_iw_jp_n\big(\theta+\tau \big(sI[Z_i=Z_j]-1\big)\big)$ and $\mathbb{E}[A_{ij}|Z]=\eta_{ij}$.  Then
\begin{eqnarray}\nonumber
&&\sum_{i\neq j\neq k}(A_{ij}-\beta_i\beta_j)(A_{jk}-\beta_j\beta_k)(d_k-\mathbb{E}[d_k])(d_i-\mathbb{E}[d_i])\\ \nonumber
&=&\sum_{i\neq j\neq k,l,t}(A_{ij}-\eta_{ij}+\eta_{ij}-\beta_i\beta_j)(A_{jk}-\eta_{jk}+\eta_{jk}-\beta_j\beta_k)\\ \nonumber
&&\times (A_{kl}-\eta_{kl}+\eta_{kl}-\beta_k\beta_l)(A_{it}-\eta_{it}+\eta_{it}-\beta_i\beta_t)\\ \nonumber
&=&\sum_{i\neq j\neq k,l,t}(A_{ij}-\eta_{ij})(A_{jk}-\eta_{jk})(A_{kl}-\eta_{kl})(A_{it}-\eta_{it})\\ \nonumber
&&+\sum_{i\neq j\neq k,l,t}(A_{ij}-\eta_{ij})(A_{jk}-\eta_{jk})(A_{kl}-\eta_{kl})(\eta_{it}-\beta_i\beta_t)\\ \nonumber
&&+\sum_{i\neq j\neq k,l,t}(A_{ij}-\eta_{ij})(A_{jk}-\eta_{jk})(\eta_{kl}-\beta_k\beta_l)(\eta_{it}-\beta_i\beta_t)\\ \nonumber
&&+\sum_{i\neq j\neq k,l,t}(A_{ij}-\eta_{ij})(\eta_{jk}-\beta_j\beta_k)(\eta_{kl}-\beta_k\beta_l)(\eta_{it}-\beta_i\beta_t)\\ \label{r1ijkeq2}
&&+\sum_{i\neq j\neq k,l,t}(\eta_{ij}-\beta_i\beta_j)(\eta_{jk}-\beta_j\beta_k)(\eta_{kl}-\beta_k\beta_l)(\eta_{it}-\beta_i\beta_t)
\end{eqnarray}

Recall that $A_{ij}$ are conditionally independent given $Z$ and $\mathbb{E}[A_{ij}|Z]=\eta_{ij}$.
It is easy to get that
\begin{eqnarray}\nonumber
&&\mathbb{E}\left[\Bigg(\sum_{\substack{i\neq j\neq k, l\neq j, t\neq j,\\ |\{t,l\}\cap\{i,k\}|\leq 1}}(A_{ij}-\eta_{ij})(A_{jk}-\eta_{jk})(A_{kl}-\eta_{kl})(A_{it}-\eta_{it})\Bigg)^2\right]\\ \nonumber
&=&\sum_{\substack{i\neq j\neq k, l\neq j, t\neq j,\\ |\{t,l\}\cap\{i,k\}|\leq 1}}\mathbb{E}\left[\mathbb{E}[(A_{ij}-\eta_{ij})^2|Z]\mathbb{E}[(A_{jk}-\eta_{jk})^2|Z]\mathbb{E}[(A_{kl}-\eta_{kl})^2|Z]\mathbb{E}[(A_{it}-\eta_{it})^2|Z]\right]\\ \label{seq1}
&=&O(n^5p_n^4),
\end{eqnarray}

\begin{eqnarray}\nonumber
&&\mathbb{E}\left[\Bigg(\sum_{\substack{i\neq j\neq k, l\neq j, t\neq j,\\ |\{t,l\}\cap\{i,k\}|=2}}(A_{ij}-\eta_{ij})(A_{jk}-\eta_{jk})(A_{ik}-\eta_{ik})^2\Bigg)^2\right]\\ \nonumber
&=&\sum_{\substack{i\neq j\neq k, l\neq j, t\neq j,\\ |\{t,l\}\cap\{i,k\}|=2}}\mathbb{E}\left[\mathbb{E}[(A_{ij}-\eta_{ij})^2|Z]\mathbb{E}[(A_{jk}-\eta_{jk})^2|Z]\mathbb{E}[(A_{ik}-\eta_{ik})^4|Z]\right]\\  \label{seq2}
&=&O(n^3p_n^3),
\end{eqnarray}

\begin{eqnarray}\nonumber
&&\mathbb{E}\left[\Bigg(\sum_{\substack{i\neq j\neq k, l=j, t\neq j,t\neq k}}(A_{ij}-\eta_{ij})(A_{jk}-\eta_{jk})^2(A_{it}-\eta_{it})\Bigg)^2\right]\\ \nonumber
&=&\sum_{\substack{i\neq j\neq k, l=j, \\ t\neq j,t\neq k,k_1\neq k}}\mathbb{E}\left[\mathbb{E}[(A_{ij}-\eta_{ij})^2|Z]\mathbb{E}[(A_{jk}-\eta_{jk})^2|Z]\mathbb{E}[(A_{jk_1}-\eta_{jk_1})^2|Z]\mathbb{E}[(A_{it}-\eta_{it})^2|Z]\right]\\ \nonumber
&&+\sum_{\substack{i\neq j\neq k, l=j, \\ t\neq j,t\neq k }}\mathbb{E}\left[\mathbb{E}[(A_{ij}-\eta_{ij})^2|Z]\mathbb{E}[(A_{jk}-\eta_{jk})^4|Z]\mathbb{E}[(A_{it}-\eta_{it})^2|Z]\right]\\ \label{seq3}
&=&O(n^5p_n^4)+O(n^4p_n^3),
\end{eqnarray}

\begin{eqnarray}\label{seq4}
\mathbb{E}\left[\Bigg(\sum_{\substack{i\neq j\neq k, l=j, t\neq j,t=k}}(A_{ij}-\eta_{ij})(A_{jk}-\eta_{jk})^2(A_{ik}-\eta_{ik})\Bigg)^2\right]=O(n^3p_n^3),
\end{eqnarray}

\begin{eqnarray}\nonumber
&&\mathbb{E}\left[\sum_{\substack{i\neq j\neq k, l=j, t=j}}(A_{ij}-\eta_{ij})^2(A_{jk}-\eta_{jk})^2\right]\\ \nonumber
&=&\sum_{\substack{i\neq j\neq k, l=j, t=j}}\mathbb{E}\left[\mathbb{E}[(A_{ij}-\eta_{ij})^2|Z]\mathbb{E}[(A_{jk}-\eta_{jk})^2|Z]\right]\\ \label{seq5}
&=&O(n^3p_n^2),
\end{eqnarray}

Combining (\ref{seq1})-(\ref{seq5}) yields
\begin{eqnarray}\label{r1ijkeq8}
\sum_{i\neq j\neq k,l,t}(A_{ij}-\eta_{ij})(A_{jk}-\eta_{jk})(A_{kl}-\eta_{kl})(A_{it}-\eta_{it})=O_P\left(n^3p_n^2\right).
\end{eqnarray}

(b). The second moment of the last term of (\ref{r1ijkeq2}) is equal to
\begin{eqnarray*}
&&\mathbb{E}\Big[\sum_{i\neq j\neq k,l,t}(\eta_{ij}-\beta_i\beta_j)(\eta_{jk}-\beta_j\beta_k)(\eta_{kl}-\beta_k\beta_l)(\eta_{it}-\beta_i\beta_t)\Big]^2\\
&=&\sum_{\substack{i\neq j\neq k,l,t\\i_1\neq j_1\neq k_1,l_1,t_1}}\mathbb{E}\Big[(\eta_{ij}-\beta_i\beta_j)(\eta_{jk}-\beta_j\beta_k)(\eta_{kl}-\beta_k\beta_l)(\eta_{it}-\beta_i\beta_t)\\
&&\times(\eta_{i_1j_1}-\beta_{i_1}\beta_{j_1})(\eta_{j_1k_1}-\beta_{j_1}\beta_{k_1})(\eta_{k_1l_1}-\beta_{k_1}\beta_{l_1})(\eta_{i_1t_1}-\beta_{i_1}\beta_{t_1})\Big].
\end{eqnarray*}
Recall that $\mathbb{E}\big[(\eta_{ij}-\beta_i\beta_j)|Z_j]=0$  by (\ref{ezeq1}). If $l_1\notin\{i\neq j\neq k,l,t,i_1,j_1,t_1\}$, then 
\begin{eqnarray*}
&&\mathbb{E}\Big[(\eta_{ij}-\beta_i\beta_j)(\eta_{jk}-\beta_j\beta_k)(\eta_{kl}-\beta_k\beta_l)(\eta_{it}-\beta_i\beta_t)\\
&&\times(\eta_{i_1j_1}-\beta_{i_1}\beta_{j_1})(\eta_{j_1k_1}-\beta_{j_1}\beta_{k_1})(\eta_{k_1l_1}-\beta_{k_1}\beta_{l_1})(\eta_{i_1t_1}-\beta_{i_1}\beta_{t_1})\Big]\\
&=&\mathbb{E}\Big[(\eta_{ij}-\beta_i\beta_j)(\eta_{jk}-\beta_j\beta_k)(\eta_{kl}-\beta_k\beta_l)(\eta_{it}-\beta_i\beta_t)\\
&&\times(\eta_{i_1j_1}-\beta_{i_1}\beta_{j_1})(\eta_{j_1k_1}-\beta_{j_1}\beta_{k_1})(\eta_{i_1t_1}-\beta_{i_1}\beta_{t_1})\\
&&\times\mathbb{E}[(\eta_{k_1l_1}-\beta_{k_1}\beta_{l_1})|Z_i,Z_j,Z_k,Z_l,Z_t,Z_{i_1},Z_{j_1},Z_{k_1},Z_{t_1}]\Big]\\
&=&0.
\end{eqnarray*}
Hence $l_1\in\{i\neq j\neq k,l,t,i_1,j_1,t_1\}$.
Similarly,  $t_1\in\{i\neq j\neq k,l,t,i_1,j_1,l_1\}$, $t\in\{i\neq j\neq k,l,i_1,j_1,k_1,t_1,l_1\}$,$l\in\{i\neq j\neq k,t,i_1,j_1,k_1,t_1,l_1\}$. There are at most $n^8$ such choices of the indices $i\neq j\neq k,l,t,i_1,j_1,k_1,l_1,t_1$. Hence, we get
\begin{eqnarray}\label{r1ijkeq5}
&&\mathbb{E}\Big[\sum_{i\neq j\neq k,l,t}(\eta_{ij}-\beta_i\beta_j)(\eta_{jk}-\beta_j\beta_k)(\eta_{kl}-\beta_k\beta_l)(\eta_{it}-\beta_i\beta_t)\Big]^2=O(n^8p_n^8).
\end{eqnarray}

(c). Consider the second moment of the second term of (\ref{r1ijkeq2}) in two cases: $l=j$ and $l\neq j$ as follows.
\begin{eqnarray}\nonumber
&&\mathbb{E}\Big[\sum_{i\neq j\neq k,l\neq j,t}(A_{ij}-\eta_{ij})(A_{jk}-\eta_{jk})(A_{kl}-\eta_{kl})(\eta_{it}-\beta_i\beta_t)\Big]^2\\ \nonumber
&=&\sum_{i\neq j\neq k,l\neq j,t,t_1}\mathbb{E}\Big[\mathbb{E}\Big[(A_{ij}-\eta_{ij})^2(A_{jk}-\eta_{jk})^2(A_{kl}-\eta_{kl})^2\big|Z\Big](\eta_{it}-\beta_i\beta_t)(\eta_{it_1}-\beta_i\beta_{t_1})\Big]\\ \label{r1ijkeq6}
&=&O(n^6p_n^5),
\end{eqnarray}
and
\begin{eqnarray}\nonumber
&&\mathbb{E}\Big[\sum_{i\neq j\neq k,l=j,t}(A_{ij}-\eta_{ij})(A_{jk}-\eta_{jk})^2(\eta_{it}-\beta_i\beta_t)\Big]^2\\ \nonumber
&=&\sum_{i\neq j\neq k,l= j,t,t_1,k_1}\mathbb{E}\Big[\mathbb{E}\Big[(A_{ij}-\eta_{ij})^2(A_{jk}-\eta_{jk})^2(A_{jk_1}-\eta_{jk_1})^2\big|Z\Big](\eta_{it}-\beta_i\beta_t)(\eta_{it_1}-\beta_i\beta_{t_1})\Big]\\ \label{r1ijkeq7}
&=&O(n^6p_n^5).
\end{eqnarray}

(d). Consider the second moment of the third term of (\ref{r1ijkeq2}).
\begin{eqnarray*}
&&\mathbb{E}\Big[\sum_{i\neq j\neq k,l,t}(A_{ij}-\eta_{ij})(A_{jk}-\eta_{jk})(\eta_{kl}-\beta_k\beta_l)(\eta_{it}-\beta_i\beta_t)\Big]^2\\
&=&\sum_{i\neq j\neq k,l,t,t_1,l_1}\mathbb{E}\Big[\mathbb{E}\Big[(A_{ij}-\eta_{ij})^2(A_{jk}-\eta_{jk})^2\big|Z\Big](\eta_{kl}-\beta_k\beta_{l})(\eta_{it}-\beta_i\beta_{t})\\
&&\times(\eta_{kl_1}-\beta_k\beta_{l_1})(\eta_{it_1}-\beta_i\beta_{t_1})\Big].
\end{eqnarray*}
If  $t\notin\{j,k,l,l_1,t_1\}$, then
\[\mathbb{E}\Big[\mathbb{E}\Big[(A_{ij}-\eta_{ij})^2(A_{jk}-\eta_{jk})^2\big|Z\Big](\eta_{kl}-\beta_k\beta_{l})(\eta_{it}-\beta_i\beta_{t})(\eta_{kl_1}-\beta_k\beta_{l_1})(\eta_{it_1}-\beta_i\beta_{t_1})\Big]=0.\]
Hence, $t\in\{j,k,l,l_1,t_1\}$. There are at most $n^6$ choices of such indices. Then
\begin{eqnarray}\label{r1ijkeq3}
\mathbb{E}\Big[\sum_{i\neq j\neq k,l,t}(A_{ij}-\eta_{ij})(A_{jk}-\eta_{jk})(\eta_{kl}-\beta_k\beta_l)(\eta_{it}-\beta_i\beta_t)\Big]^2=O(n^6p_n^6).
\end{eqnarray}

(e). The second moment of the fourth term of (\ref{r1ijkeq2}) is equal to
\begin{eqnarray*}
&&\mathbb{E}\Big[\sum_{i\neq j\neq k,l,t}(A_{ij}-\eta_{ij})(\eta_{jk}-\beta_j\beta_k)(\eta_{kl}-\beta_k\beta_l)(\eta_{it}-\beta_i\beta_t)\Big]^2\\
&=&\sum_{i\neq j\neq k,l,t,k_1,t_1,l_1}\mathbb{E}\Big[\mathbb{E}\Big[(A_{ij}-\eta_{ij})^2\big|Z\Big](\eta_{jk}-\beta_j\beta_{k})(\eta_{kl}-\beta_k\beta_{l})(\eta_{it}-\beta_i\beta_{t})\\
&&\times(\eta_{jk_1}-\beta_j\beta_{k_1})(\eta_{k_1l_1}-\beta_{k_1}\beta_{l_1})(\eta_{it_1}-\beta_i\beta_{t_1})\Big].
\end{eqnarray*}
By a similar argument as in (\ref{r1ijkeq3}), we have $l\in\{i\neq j\neq k,t,k_1,t_1,l_1\}$, $t\in\{i\neq j\neq k,l,k_1,t_1,l_1\}$, $l_1\in\{i\neq j\neq k,l,t,k_1,t_1\}$ and $t_1\in\{i\neq j\neq k,t,k_1,l_1\}$. There are at most $n^6$ choices of such indices. 
Hence
\begin{eqnarray}\label{r1ijkeq4}
\mathbb{E}\Big[\sum_{i\neq j\neq k,l,t}(A_{ij}-\eta_{ij})(\eta_{jk}-\beta_j\beta_k)(\eta_{kl}-\beta_k\beta_l)(\eta_{it}-\beta_i\beta_t)\Big]^2=O(n^6p_n^6).
\end{eqnarray}

By Lemma \ref{plem1}, $d=\mu+o_P(\mu)$. 
Combining (\ref{r1ijkeq8})-(\ref{r1ijkeq4}) and Markov's inequality, we have
\begin{eqnarray}\nonumber
&&\frac{1}{d}\sum_{i\neq j\neq k}(A_{ij}-\beta_i\beta_j)(A_{jk}-\beta_j\beta_k)(d_k-\mathbb{E}[d_k])(d_i-\mathbb{E}[d_i])\\ \label{r1ijkeq9}
&&-\frac{1}{d}\sum_{i\neq j\neq k,l,t}(\eta_{ij}-\beta_i\beta_j)(\eta_{jk}-\beta_j\beta_k)(\eta_{kl}-\beta_k\beta_l)(\eta_{it}-\beta_i\beta_t)=o_P\left(\sqrt{(np_n)^3}\right).
\end{eqnarray}

The second term of (\ref{r1ijkeq1}) is equal to
\begin{eqnarray}\nonumber
&&\sum_{i\neq j\neq k}(A_{ij}-\beta_i\beta_j)(A_{jk}-\beta_j\beta_k)(d_k-\mathbb{E}[d_k])\mu_i\\ \nonumber
&=&\sum_{i\neq j\neq k,l}(A_{ij}-\eta_{ij}+\eta_{ij}-\beta_i\beta_j)(A_{jk}-\eta_{jk}+\eta_{jk}-\beta_j\beta_k)(A_{kl}-\eta_{kl}+\eta_{kl}-\beta_k\beta_l)\mu_i\\ \nonumber
&=&\sum_{i\neq j\neq k,l}(A_{ij}-\eta_{ij})(A_{jk}-\eta_{jk})(A_{kl}-\eta_{kl})\mu_i\\ \nonumber
&&+\sum_{i\neq j\neq k,l}(A_{ij}-\eta_{ij})(A_{jk}-\eta_{jk})(\eta_{kl}-\beta_k\beta_l)\mu_i\\ \nonumber
&&+\sum_{i\neq j\neq k,l}(A_{ij}-\eta_{ij})(\eta_{jk}-\beta_j\beta_k)(\eta_{kl}-\beta_k\beta_l)\mu_i\\ \label{r1ijkeq10}
&&+\sum_{i\neq j\neq k,l}(\eta_{ij}-\beta_i\beta_j)(\eta_{jk}-\beta_j\beta_k)(\eta_{kl}-\beta_k\beta_l)\mu_i.
\end{eqnarray}
Similar to (\ref{r1ijkeq8}), it is easy to verify that
\begin{eqnarray}\label{r1ijkeq11}
&&\sum_{i\neq j\neq k,l}(A_{ij}-\eta_{ij})(A_{jk}-\eta_{jk})(A_{kl}-\eta_{kl})\mu_i=O_P\left(\sqrt{n^6p_n^5}\right).
\end{eqnarray}
The second moment of the second term of (\ref{r1ijkeq10}) is bounded by
\begin{eqnarray}\nonumber
&&\mathbb{E}\Big[\sum_{i\neq j\neq k,l}(A_{ij}-\eta_{ij})(A_{jk}-\eta_{jk})(\eta_{kl}-\beta_k\beta_l)\mu_i\Big]^2\\ \nonumber
&=&\sum_{i\neq j\neq k,l,l_1}\mathbb{E}\Big[\mathbb{E}\Big[(A_{ij}-\eta_{ij})^2(A_{jk}-\eta_{jk})^2\big|Z\Big](\eta_{kl}-\beta_k\beta_l)(\eta_{kl_1}-\beta_k\beta_{l_1})\mu_i^2\Big]\\ \label{r1ijkeq12}
&=&O\left(n^7p_n^6\right).
\end{eqnarray}
The second moment of the third term of (\ref{r1ijkeq10}) is bounded by
\begin{eqnarray}\nonumber
&&\mathbb{E}\Big[\sum_{i\neq j\neq k,l}(A_{ij}-\eta_{ij})(\eta_{jk}-\beta_j\beta_k)(\eta_{kl}-\beta_k\beta_l)\mu_i\Big]^2\\ \nonumber
&=&\sum_{i\neq j\neq k,l,k_1,l_1}\mathbb{E}\Big[\mathbb{E}\Big[(A_{ij}-\eta_{ij})^2|Z\Big](\eta_{jk}-\beta_j\beta_k)(\eta_{kl}-\beta_k\beta_l)\mu_i^2\\ \nonumber
&&\times (\eta_{jk_1}-\beta_j\beta_{k_1})(\eta_{k_1l_1}-\beta_{k_1}\beta_{l_1})\Big]\\ \label{r1ijkeq13}
&=&O(n^8p_n^7).
\end{eqnarray}
The last term of (\ref{r1ijkeq10}) is bounded by
\begin{eqnarray}\label{r1ijkeq14}
\sum_{i\neq j\neq k,l}(\eta_{ij}-\beta_i\beta_j)(\eta_{jk}-\beta_j\beta_k)(\eta_{kl}-\beta_k\beta_l)\mu_i=O_P(n^4p_n^4).
\end{eqnarray}

Combining (\ref{r1ijkeq10})-(\ref{r1ijkeq14}), Markov's inequality and Lemma \ref{plem1} yields
\begin{eqnarray}\label{r1ijkeq15}
\frac{2(\sqrt{d}-\sqrt{\mu})}{d\sqrt{\mu}}\sum_{i\neq j\neq k}(A_{ij}-\beta_i\beta_j)(A_{jk}-\beta_j\beta_k)\mu_i(d_k-\mathbb{E}[d_k])=o_P(\sqrt{(np_n)^3}).
\end{eqnarray}

Now consider the last term of (\ref{r1ijkeq1}).
\begin{eqnarray}\nonumber
&&\sum_{i\neq j\neq k}(A_{ij}-\beta_i\beta_j)(A_{jk}-\beta_j\beta_k)\mu_k\mu_i\\ \nonumber
&=&\sum_{i\neq j\neq k}(A_{ij}-\eta_{ij}+\eta_{ij}-\beta_i\beta_j)(A_{jk}-\eta_{jk}+\eta_{jk}-\beta_j\beta_k)\mu_k\mu_i\\ \nonumber
&=&\sum_{i\neq j\neq k}(A_{ij}-\eta_{ij})(A_{jk}-\eta_{jk})\mu_k\mu_i+\sum_{i\neq j\neq k}(A_{ij}-\eta_{ij})(\eta_{jk}-\beta_j\beta_k)\mu_k\mu_i\\ \label{r1ijkeq16}
&&+\sum_{i\neq j\neq k}(\eta_{ij}-\beta_i\beta_j)(\eta_{jk}-\beta_j\beta_k)\mu_k\mu_i
\end{eqnarray}

It is easy to verify that
\begin{eqnarray}\label{r1ijkeq17}
\sum_{i\neq j\neq k}(A_{ij}-\eta_{ij})(A_{jk}-\eta_{jk})\mu_k\mu_i&=&O_P(n^2p_n^2\sqrt{n^3p_n^2}),\\ \label{r1ijkeq17}
\sum_{i\neq j\neq k}(A_{ij}-\eta_{ij})(\eta_{jk}-\beta_j\beta_k)\mu_k\mu_i&=&O_P(n^2p_n^2\sqrt{n^4p_n^3}),\\ \label{r1ijkeq18}
\sum_{i\neq j\neq k}(\eta_{ij}-\beta_i\beta_j)(\eta_{jk}-\beta_j\beta_k)\mu_k\mu_i&=&O_P\left(n^5p_n^4\right).
\end{eqnarray}

Combining (\ref{r1ijkeq1})-(\ref{r1ijkeq18}) yields
\begin{eqnarray}\label{r1ijkeq19}
\sum_{i\neq j\neq k}(A_{ij}-\beta_i\beta_j)(A_{jk}-\beta_j\beta_k)(b_k-\beta_k)(b_i-\beta_i)=o_P(\sqrt{n^3p_n^3}).
\end{eqnarray}

By the proof of the first term of (\ref{r1ijkeq19}), it is easy to get
\begin{eqnarray}\nonumber
&&\sum_{i\neq j\neq k}(A_{ij}-\beta_i\beta_j)(A_{jk}-\beta_j\beta_k)(b_k-\beta_k)\beta_i\\ \nonumber
&=&\sum_{i\neq j\neq k}(A_{ij}-\beta_i\beta_j)(A_{jk}-\beta_j\beta_k)\left(\frac{d_k-\mu_k}{\sqrt{d}}-\frac{(\sqrt{d}-\sqrt{\mu})\mu_k}{\sqrt{d\mu}}\right)\beta_i\\ \nonumber
&=&\frac{1}{\sqrt{d\mu}}\sum_{i\neq j\neq k}(A_{ij}-\beta_i\beta_j)(A_{jk}-\beta_j\beta_k)(d_k-\mu_k)\mu_i\\ \nonumber
&&-\frac{(\sqrt{d}-\sqrt{\mu})}{\mu\sqrt{d}}\sum_{i\neq j\neq k}(A_{ij}-\beta_i\beta_j)(A_{jk}-\beta_j\beta_k)\mu_k\mu_i\\ \label{r1ijkeq20}
&=&o_P(\sqrt{n^3p_n^3}).
\end{eqnarray}

Combining (\ref{r1ijkeq19}) and (\ref{r1ijkeq20}) yields $\sum_{i\neq j\neq k}R_{1,i\neq j\neq k}=o_P(np_n\sqrt{np_n})$.

\vskip 1cm

{\bf Step 2:} we prove $\sum_{i\neq j\neq k}R_{2,ijk}=o_P(np_n\sqrt{np_n})$. We shall bound the summation of each term in $R_{2,ijk}$ over $i\neq j\neq k$. 

Straightforward calculation yields
\begin{eqnarray}\nonumber
&&\sum_{i\neq j\neq k}(A_{ij}-\beta_i\beta_j)(b_j-\beta_j)(b_k-\beta_k)^2(b_i-\beta_i)\\ \nonumber
&=&\sum_{i\neq j\neq k}(A_{ij}-\beta_i\beta_j)\left(\frac{d_i-\mu_i}{\sqrt{d}}-\frac{(\sqrt{d}-\sqrt{\mu})\mu_i}{\sqrt{d\mu}}\right)\left(\frac{d_j-\mu_j}{\sqrt{d}}-\frac{(\sqrt{d}-\sqrt{\mu})\mu_j}{\sqrt{d\mu}}\right)\\ \nonumber
&&\times\left(\frac{d_k-\mu_k}{\sqrt{d}}-\frac{(\sqrt{d}-\sqrt{\mu})\mu_k}{\sqrt{d\mu}}\right)^2\\ \nonumber
&=&\frac{1}{d^2}\sum_{i\neq j\neq k}(A_{ij}-\beta_i\beta_j)(d_i-\mu_i)(d_j-\mu_j)(d_k-\mu_k)^2\\ \nonumber
&&-\frac{2(\sqrt{d}-\sqrt{\mu})}{d^2\sqrt{\mu}}\sum_{i\neq j\neq k}(A_{ij}-\beta_i\beta_j)(d_i-\mu_i)(d_j-\mu_j)(d_k-\mu_k)\mu_k\\ \nonumber
&&+\frac{(\sqrt{d}-\sqrt{\mu})^2}{d^2\mu}\sum_{i\neq j\neq k}(A_{ij}-\beta_i\beta_j)(d_i-\mu_i)(d_j-\mu_j)\mu_k^2\\ \nonumber
&&-2\frac{(\sqrt{d}-\sqrt{\mu})}{d^2\sqrt{\mu}}\sum_{i\neq j\neq k}(A_{ij}-\beta_i\beta_j)(d_i-\mu_i)(d_k-\mu_k)^2\mu_j\\ \nonumber
&&+4\frac{(\sqrt{d}-\sqrt{\mu})^2}{d^2\mu}\sum_{i\neq j\neq k}(A_{ij}-\beta_i\beta_j)(d_i-\mu_i)(d_k-\mu_k)\mu_j\mu_k\\ \nonumber
&&-2\frac{(\sqrt{d}-\sqrt{\mu})^3}{d^2\mu\sqrt{\mu}}\sum_{i\neq j\neq k}(A_{ij}-\beta_i\beta_j)(d_i-\mu_i)\mu_j\mu_k^2\\ \nonumber
&&+\frac{(\sqrt{d}-\sqrt{\mu})^2}{d^2\mu}\sum_{i\neq j\neq k}(A_{ij}-\beta_i\beta_j)(d_k-\mu_k)^2\mu_j\mu_i\\ \nonumber
&&-2\frac{(\sqrt{d}-\sqrt{\mu})^3}{d^2\mu\sqrt{\mu}}\sum_{i\neq j\neq k}(A_{ij}-\beta_i\beta_j)(d_k-\mu_k)\mu_i\mu_j\mu_k\\ \label{sueq1}
&&+\frac{(\sqrt{d}-\sqrt{\mu})^4}{d^2\mu^2}\sum_{i\neq j\neq k}(A_{ij}-\beta_i\beta_j)\mu_i\mu_j\mu_k^2.
\end{eqnarray}

Next we bound each term in (\ref{sueq1}). The first term of (\ref{sueq1}) can be expressed as
\begin{eqnarray}\nonumber
&&\sum_{i\neq j\neq k}(A_{ij}-\beta_i\beta_j)(d_i-\mu_i)(d_j-\mu_j)(d_k-\mu_k)^2\\ \nonumber
&=&\sum_{i\neq j}(A_{ij}-\beta_i\beta_j)(d_i-\mu_i)(d_j-\mu_j)\Big[\sum_k(d_k-\mu_k)^2-(d_i-\mu_i)^2-(d_j-\mu_j)^2\Big]\\  \nonumber
&=&\left(\sum_{i\neq j}(A_{ij}-\beta_i\beta_j)(d_i-\mu_i)(d_j-\mu_j)\right)\left(\sum_k(d_k-\mu_k)^2\right)\\ \label{sueq2}
&&-2\sum_{i\neq j}(A_{ij}-\beta_i\beta_j)(d_i-\mu_i)^3(d_j-\mu_j).
\end{eqnarray}

By a similar argument as the proof of Lemma \ref{lemeta}, it is easy to get
\begin{eqnarray}\nonumber
\sum_{i\neq j}(A_{ij}-\beta_i\beta_j)(d_i-\mu_i)(d_j-\mu_j)&=&\sum_{i\neq j\neq k\neq l}(A_{ij}-\beta_i\beta_j)(A_{ik}-\beta_i\beta_k)(A_{jl}-\beta_j\beta_l)\\ \nonumber
&&+\sum_{i\neq j\neq k}(A_{ij}-\beta_i\beta_j)(A_{ik}-\beta_i\beta_k)(A_{jk}-\beta_j\beta_k)\\ \nonumber
&&+\sum_{i\neq j\neq k}(A_{ij}-\beta_i\beta_j)^2(A_{ik}-\beta_i\beta_k)\\ \label{sueq3}
&=&O_P\left(n^3p_n^3\right).
\end{eqnarray}

By Lemma \ref{lemdmu}, we have
\begin{eqnarray}\label{sueq4}
\sum_k(d_k-\mu_k)^2=O_P\left(n^2p_n\right).
\end{eqnarray}

By the Cauchy-Schwarz inequality and Lemma \ref{lemdmu}, we have
\begin{eqnarray}\nonumber
&&\sum_{i,j}\mathbb{E}\Big[|A_{ij}-\beta_i\beta_j||d_i-\mu_i|^3|d_j-\mu_j|\Big]\\ \nonumber
&\leq&\sum_{i,j}\sqrt{\mathbb{E}\big[(A_{ij}-\beta_i\beta_j)^2\big]\sqrt{\mathbb{E}\Big[(d_i-\mu_i)^{12}\Big]\mathbb{E}\Big[(d_j-\mu_j)^4\Big]}}\\ \label{sueq5}
&=&O\left(n^2(np_n)^2\sqrt{p_n}\right).
\end{eqnarray}
Combining (\ref{sueq2})-(\ref{sueq5}) and Lemma \ref{plem1} yields
\begin{eqnarray*}
\frac{1}{d^2}\sum_{i\neq j\neq k}(A_{ij}-\beta_i\beta_j)(d_i-\mu_i)(d_j-\mu_j)(d_k-\mu_k)^2=o_P\left(\sqrt{(np_n)^3}\right).
\end{eqnarray*}

The second term of (\ref{sueq1}) can be expressed as
\begin{eqnarray}\nonumber
&&\sum_{i\neq j\neq k}(A_{ij}-\beta_i\beta_j)(d_i-\mu_i)(d_j-\mu_j)(d_k-\mu_k)\mu_k\\ \nonumber
&=&\left(\sum_{i\neq j}(A_{ij}-\beta_i\beta_j)(d_i-\mu_i)(d_j-\mu_j)\right)\left(\sum_k(d_k-\mu_k)\mu_k\right)\\ \label{sueq5}
&&-2\sum_{i\neq j}(A_{ij}-\beta_i\beta_j)(d_i-\mu_i)^2(d_j-\mu_j)\mu_i.
\end{eqnarray}
The first term of (\ref{sueq5}) can be bounded by (\ref{sueq3}) and using Lemma \ref{lemdmu}. We only need to bound the second term of (\ref{sueq5}).  By the Cauchy–Schwarz inequality, we have
\begin{eqnarray}\nonumber
&&\sum_{i\neq j}\mathbb{E}\Big[\big|(A_{ij}-\beta_i\beta_j)(d_i-\mu_i)^2(d_j-\mu_j)\big|\Big]\mu_i\\ \label{sueq6}
&\leq &\sum_{i\neq j}\mu_i\sqrt{\mathbb{E}\big[(A_{ij}-\beta_i\beta_j)^2(d_j-\mu_j)^2\big]}\sqrt{\mathbb{E}(d_i-\mu_i)^4}.
\end{eqnarray}
For given distinct indices $i,j$, it is easy to verify that
\begin{eqnarray}\nonumber
&&\mathbb{E}\big[(A_{ij}-\beta_i\beta_j)^2(d_j-\mu_j)^2\big]\\ \nonumber
&=&
\mathbb{E}\Big[(A_{ij}-\beta_i\beta_j)^2\sum_{k\neq s}(A_{jk}-\beta_j\beta_k)(A_{js}-\beta_j\beta_s)\Big]+\mathbb{E}\Big[(A_{ij}-\beta_i\beta_j)^2\sum_{k\neq i}(A_{jk}-\beta_j\beta_k)^2\Big]\\ \label{sbmeq1}
&&+\mathbb{E}\Big[(A_{ij}-\beta_i\beta_j)^4\Big].
\end{eqnarray}
Next we show the first term in (\ref{sbmeq1}) vanishes. Straightforward calculation yields
\begin{eqnarray} \nonumber
&&(A_{ij}-\beta_i\beta_j)^2(A_{jk}-\beta_j\beta_k)(A_{js}-\beta_j\beta_s)\\ \nonumber
&=&(A_{ij}-\eta_{ij}+\eta_{ij}-\beta_i\beta_j)^2(A_{jk}-\eta_{jk}+\eta_{jk}-\beta_j\beta_k)(A_{js}-\eta_{js}+\eta_{js}-\beta_j\beta_s)\\  \nonumber
&=&\Big[(A_{ij}-\eta_{ij})^2+(\eta_{ij}-\beta_i\beta_j)^2+2(A_{ij}-\eta_{ij})(\eta_{ij}-\beta_i\beta_j)\Big]\\  \nonumber
&&\times\Big[(A_{jk}-\eta_{jk})(A_{js}-\eta_{js})+(A_{jk}-\eta_{jk})(\eta_{js}-\beta_j\beta_s)+(\eta_{jk}-\beta_j\beta_k)(A_{js}-\eta_{js})\\ \label{sueq7}
&&+(\eta_{jk}-\beta_j\beta_k)(\eta_{js}-\beta_j\beta_s)\Big].
\end{eqnarray}
Note that $k\neq s$. Assume  $k\neq i$ or $s\neq i$. Let $s\neq i$ without loss of generality. Then
\begin{eqnarray}\label{sueq8}
\mathbb{E}\Big[(A_{ij}-\eta_{ij})^2(A_{jk}-\eta_{jk})(A_{js}-\eta_{js})\Big]=\mathbb{E}\Big[(A_{ij}-\eta_{ij})^2(A_{jk}-\eta_{jk})\mathbb{E}\big[(A_{js}-\eta_{js})\big|Z\big]\Big]=0.
\end{eqnarray}

For $k\neq i$,  
\begin{eqnarray}\label{sueq9}
\mathbb{E}\Big[(A_{ij}-\eta_{ij})^2(A_{jk}-\eta_{jk})(\eta_{js}-\beta_j\beta_s)\Big]=\mathbb{E}\Big[(A_{ij}-\eta_{ij})^2\mathbb{E}\big[(A_{jk}-\eta_{jk})\big|Z\big](\eta_{js}-\beta_j\beta_s)\Big]=0.
\end{eqnarray}
For $k=i$, then 
\begin{eqnarray}\label{sueq10}
\mathbb{E}\Big[(A_{ij}-\eta_{ij})^3(\eta_{js}-\beta_j\beta_s)\Big]=\mathbb{E}\Big[\mathbb{E}\big[(A_{ij}-\eta_{ij})^3\big|Z\big]\mathbb{E}\big[(\eta_{js}-\beta_j\beta_s)|Z_i,Z_j\big]\Big]=0.
\end{eqnarray}

Suppose $s\neq i$. Then
\begin{eqnarray} \nonumber
&&\mathbb{E}\Big[(A_{ij}-\eta_{ij})^2(\eta_{jk}-\beta_j\beta_k)(\eta_{js}-\beta_j\beta_s)\Big]\\  \nonumber
&=&\mathbb{E}\Big[\mathbb{E}\big[(A_{ij}-\eta_{ij})^2\big|Z\big](\eta_{jk}-\beta_j\beta_k)(\eta_{js}-\beta_j\beta_s)\Big]\\  \nonumber
&=&\mathbb{E}\Big[\mathbb{E}\big[(A_{ij}-\eta_{ij})^2\big|Z\big](\eta_{jk}-\beta_j\beta_k)\mathbb{E}\Big[(\eta_{js}-\beta_j\beta_s)\big|Z_i,Z_j,Z_k\Big]\Big]\\ \label{sueq11}
&=&0.
\end{eqnarray}

Similarly, it is easy to verify the following equations:
\begin{eqnarray*}
\mathbb{E}\Big[(\eta_{ij}-\beta_i\beta_j)^2(A_{jk}-\eta_{jk})(A_{js}-\eta_{js})\Big]&=&0,\\
\mathbb{E}\Big[(\eta_{ij}-\beta_i\beta_j)^2(A_{jk}-\eta_{jk})(\eta_{js}-\beta_j\beta_s)\Big]&=&0,\\
\mathbb{E}\Big[(\eta_{ij}-\beta_i\beta_j)^2(\eta_{jk}-\beta_j\beta_k)(\eta_{js}-\beta_j\beta_s)\Big]&=&0,\\
\mathbb{E}\Big[(A_{ij}-\eta_{ij})(\eta_{ij}-\beta_i\beta_j)(A_{jk}-\eta_{jk})(A_{js}-\eta_{js})\Big]&=&0,\\
\mathbb{E}\Big[(A_{ij}-\eta_{ij})(\eta_{ij}-\beta_i\beta_j)(A_{jk}-\eta_{jk})(\eta_{js}-\beta_j\beta_s)\Big]&=&0,\\
\mathbb{E}\Big[(A_{ij}-\eta_{ij})(\eta_{ij}-\beta_i\beta_j)(\eta_{jk}-\beta_j\beta_k)(\eta_{js}-\beta_j\beta_s)\Big]&=&0,\\
\mathbb{E}\Big[(A_{ij}-\eta_{ij})(\eta_{ij}-\beta_i\beta_j)(\eta_{jk}-\beta_j\beta_k)(A_{js}-\eta_{js})\Big]&=&0.
\end{eqnarray*}

Hence the first term in (\ref{sbmeq1}) vanishes. 

The second term in (\ref{sbmeq1}) can be bounded by
\begin{eqnarray*}
&&\mathbb{E}\Big[(A_{ij}-\beta_i\beta_j)^2\sum_{k\neq i}(A_{jk}-\beta_j\beta_k)^2\Big]\\
&\leq& 4\sum_{k\neq i}\mathbb{E}\Big[\big[(A_{ij}-\eta_{ij})^2+(\eta_{ij}-\beta_i\beta_j)^2\big]\big[(A_{jk}-\eta_{jk})^2+(\eta_{jk}-\beta_j\beta_k)^2\big]\Big]\\
&=&O(np_n^2).
\end{eqnarray*}

Hence, we get
\begin{eqnarray}\nonumber
\mathbb{E}\big[(A_{ij}-\beta_i\beta_j)^2(d_j-\mu_j)^2\big]=
O(np_n^2)
\end{eqnarray}
uniformly for all $i,j$.

Then by (\ref{sueq6}) and Lemma \ref{lemdmu}, we have
\begin{eqnarray*}
\sum_{i\neq j}\mathbb{E}\Big[\big|(A_{ij}-\beta_i\beta_j)(d_i-\mu_i)^2(d_j-\mu_j)\big|\Big]\mu_i=O(n^4p_n^3\sqrt{n}).
\end{eqnarray*}

Hence
\begin{eqnarray*}
\sum_{i\neq j}(A_{ij}-\beta_i\beta_j)(d_i-\mu_i)^2(d_j-\mu_j)\mu_i=O_P(n^4p_n^3\sqrt{n})
\end{eqnarray*}

Then
\begin{eqnarray}\label{ssseq2}
\frac{2(\sqrt{d}-\sqrt{\mu})}{d^2\sqrt{\mu}}\sum_{i\neq j\neq k}(A_{ij}-\beta_i\beta_j)(d_i-\mu_i)(d_j-\mu_j)(d_k-\mu_k)\mu_k=o_P\left(\sqrt{n^3p_n^3}\right).
\end{eqnarray}

The other terms in (\ref{sueq1}) can be similarly bounded. Then 
\begin{eqnarray*}
\sum_{i\neq j\neq k}(A_{ij}-\beta_i\beta_j)(b_j-\beta_j)(b_k-\beta_k)^2(b_i-\beta_i)=o_P\left(\sqrt{n^3p_n^3}\right).
\end{eqnarray*}

Now we consider the sum of the third term of $R_{2,ijk}$. Direct calculation yields
\begin{eqnarray}\nonumber
&&\sum_{i\neq j\neq k}(A_{ij}-\beta_i\beta_j)(b_i-\beta_i)(b_j-\beta_j)(b_k-\beta_k)\beta_k\\ \nonumber
&=&\sum_{i\neq j\neq k}(A_{ij}-\beta_i\beta_j)
\left(\frac{d_i-\mu_i}{\sqrt{d}}-\frac{(\sqrt{d}-\sqrt{\mu})\mu_i}{\sqrt{d\mu}}\right)\left(\frac{d_j-\mu_j}{\sqrt{d}}-\frac{(\sqrt{d}-\sqrt{\mu})\mu_j}{\sqrt{d\mu}}\right)\\ \nonumber
&&\times\left(\frac{d_k-\mu_k}{\sqrt{d}}-\frac{(\sqrt{d}-\sqrt{\mu})\mu_k}{\sqrt{d\mu}}\right)\mu_k\\ \nonumber
&=&\frac{1}{d\sqrt{d}}\sum_{i\neq j\neq k}(A_{ij}-\beta_i\beta_j)(d_i-\mu_i)(d_j-\mu_j)(d_k-\mu_k)\mu_k\\ \nonumber
&&-\frac{\sqrt{d}-\sqrt{\mu}}{d\sqrt{d\mu}}\sum_{i\neq j\neq k}(A_{ij}-\beta_i\beta_j)(d_i-\mu_i)(d_j-\mu_j)\mu_k^2\\ \nonumber
&&-2\frac{\sqrt{d}-\sqrt{\mu}}{d\sqrt{d\mu}}\sum_{i\neq j\neq k}(A_{ij}-\beta_i\beta_j)(d_i-\mu_i)(d_k-\mu_k)\mu_j\mu_k\\ \nonumber
&&+2\frac{(\sqrt{d}-\sqrt{\mu})^2}{d\mu\sqrt{d}}\sum_{i\neq j\neq k}(A_{ij}-\beta_i\beta_j)(d_i-\mu_i)\mu_j\mu_k^2\\ \nonumber
&&+\frac{(\sqrt{d}-\sqrt{\mu})^2}{d\mu\sqrt{d}}\sum_{i\neq j\neq k}(A_{ij}-\beta_i\beta_j)(d_k-\mu_k)\mu_i\mu_j\mu_k\\ \label{ssseq1}
&&-\frac{(\sqrt{d}-\sqrt{\mu})^3}{d\mu\sqrt{d\mu}}\sum_{i\neq j\neq k}(A_{ij}-\beta_i\beta_j)\mu_i\mu_j\mu_k^2.
\end{eqnarray}

The first term of (\ref{ssseq1}) is similarly bounded as in (\ref{ssseq2}). The second term of (\ref{ssseq1}) can be similarly bounded as in (\ref{sueq3}).
We bound the remaining terms below.

Since
\begin{eqnarray}\label{ssseq3}
\sum_{i\neq j\neq k}\mathbb{E}\big[|A_{ij}-\beta_i\beta_j|\mu_i\mu_j\mu_k^2\big]=O\left(n^3p_n(np_n)^4\right),
\end{eqnarray}
the last term of (\ref{ssseq1}) is bounded by
\begin{eqnarray}\label{ssseq33}
\frac{(\sqrt{d}-\sqrt{\mu})^3}{d\mu\sqrt{d\mu}}\sum_{i\neq j\neq k}(A_{ij}-\beta_i\beta_j)\mu_i\mu_j\mu_k^2=O\left(np_n\right).
\end{eqnarray}

The second last term of (\ref{ssseq1}) can be expressed as
\begin{eqnarray}\nonumber
&&\sum_{i\neq j\neq k}(A_{ij}-\beta_i\beta_j)(d_k-\mu_k)\mu_i\mu_j\mu_k\\ \nonumber
&=&\sum_{i\neq j}(A_{ij}-\beta_i\beta_j)\mu_i\mu_j\Big[\sum_{k}(d_k-\mu_k)\mu_k-(d_i-\mu_i)\mu_i-(d_j-\mu_j)\mu_j\Big]\\ \label{ssseq4}
&=&\left(\sum_{i\neq j}(A_{ij}-\beta_i\beta_j)\mu_i\mu_j\right)\left(\sum_{k}(d_k-\mu_k)\mu_k\right)-2\sum_{i\neq j}(A_{ij}-\beta_i\beta_j)(d_i-\mu_i)\mu_i^2\mu_j.
\end{eqnarray}
It is easy to verify that
\begin{eqnarray}\label{ssseq5}
\sum_{k}\mathbb{E}[|d_k-\mu_k|]\mu_k=O\left(n(np_n)\sqrt{np_n}\right),
\end{eqnarray}
and
\begin{eqnarray}\label{ssseq6}
\sum_{i\neq j}(A_{ij}-\beta_i\beta_j)\mu_i\mu_j=O_P\left(n^2p_n^2\sqrt{n^2p_n}\right).
\end{eqnarray}
By (\ref{r1ijkeq16}) and (\ref{ssseq6}), the last term of (\ref{ssseq4}) can be bounded as
\begin{eqnarray}\nonumber
&&\sum_{i\neq j}(A_{ij}-\beta_i\beta_j)(d_i-\mu_i)\mu_i^2\mu_j\\  \nonumber
&=&\sum_{i\neq j\neq k}(A_{ij}-\beta_i\beta_j)(A_{ik}-\beta_i\beta_k)\mu_i^2\mu_j+\sum_{i\neq j}(A_{ij}-\beta_i\beta_j)^2\mu_i^2\mu_j\\ \label{ssseq7}
&=&O(n^6p_n^5).
\end{eqnarray}

Then 
\begin{eqnarray}\label{ssseq8}
\frac{(\sqrt{d}-\sqrt{\mu})^2}{d\mu\sqrt{d}}\sum_{i\neq j\neq k}(A_{ij}-\beta_i\beta_j)(d_k-\mu_k)\mu_i\mu_j\mu_k=O_P\left(\sqrt{np_n}\right).
\end{eqnarray}

The fourth term of (\ref{ssseq1}) can be similarly bounded  as (\ref{ssseq8}). The third term of (\ref{ssseq1}) can be expressed as
\begin{eqnarray*}
&&\sum_{i\neq j\neq k}(A_{ij}-\beta_i\beta_j)(d_i-\mu_i)(d_k-\mu_k)\mu_j\mu_k\\
&=&\Big(\sum_{i\neq j}(A_{ij}-\beta_i\beta_j)(d_i-\mu_i)\mu_j\Big)\Big(\sum_k(d_k-\mu_k)\mu_k\Big)-\sum_{i,j}(A_{ij}-\beta_i\beta_j)(d_i-\mu_i)^2\mu_i\mu_j\\
&&-\sum_{i\neq j}(A_{ij}-\beta_i\beta_j)(d_i-\mu_i)(d_j-\mu_j)\mu_j^2\Big).
\end{eqnarray*}
It is easy to verify that
\begin{eqnarray*}
&&\sum_{i\neq j}(A_{ij}-\beta_i\beta_j)(d_i-\mu_i)^2\mu_i\mu_j\\
&=&\sum_{i\neq j\neq k,l}(A_{ij}-\beta_i\beta_j)(A_{ik}-\beta_i\beta_k)(A_{il}-\beta_i\beta_l)\mu_i\mu_j\\
&&+\sum_{i\neq j\neq k}(A_{ij}-\beta_i\beta_j)(A_{ik}-\beta_i\beta_k)^2\mu_i\mu_j+\sum_{i\neq j}(A_{ij}-\beta_i\beta_j)^3\mu_i\mu_j\\
&=&O_P\left(\sqrt{n(np_n)^7}\right).
\end{eqnarray*}
Then
\begin{eqnarray*}
\frac{\sqrt{d}-\sqrt{\mu}}{d\sqrt{d\mu}}\sum_{i\neq j\neq k}(A_{ij}-\beta_i\beta_j)(d_i-\mu_i)(d_k-\mu_k)\mu_j\mu_k=O_P(\sqrt{(np_n)^3}).
\end{eqnarray*}

\medskip

{\bf Step 3: } we prove $\sum_{i\neq j\neq k}R_{3,ijk}=o_P(np_n\sqrt{np_n})$. By Lemma \ref{prlem}, we have
\begin{eqnarray*}
\sum_{i\neq j\neq k}(b_i-\beta_i)^2(b_j-\beta_j)^2(b_k-\beta_k)^2
\leq \left(\sum_{i}(b_i-\beta_i)^2\right)^3=O_P\left(1\right).
\end{eqnarray*}

Note that $\beta_i=O(\sqrt{p_n})$ uniformly for all $i$. Moreover, it is easy to verify that
\begin{eqnarray*}
\sum_{i}
(b_i-\beta_i)\beta_i&=&\sum_{i}\frac{d_i-\mu_i}{\sqrt{d}}\beta_i-\sum_{i}\frac{(\sqrt{d}-\sqrt{\mu})\mu_i}{\sqrt{d\mu}}\beta_i\\
&=&\frac{1}{\sqrt{d}}\sum_{i\neq j}(A_{ij}-\beta_i\beta_j)\beta_i-\frac{(\sqrt{d}-\sqrt{\mu})}{\sqrt{d\mu}}\sum_{i}\beta_i\mu_i\\
&=&O_P\left(1\right).
\end{eqnarray*}
Then
\begin{eqnarray*}
&&\sum_{i\neq j\neq k}(b_i-\beta_i)(b_j-\beta_j)^2(b_k-\beta_k)^2\beta_i\\
&=&\left(\sum_{i}(b_i-\beta_i)\beta_i\right)\left(\sum_{i}(b_i-\beta_i)^2\right)\left(\sum_{k}(b_k-\beta_k)^2\right)\\
&&-c_1\left(\sum_{i}(b_i-\beta_i)^3\beta_i\right)\left(\sum_{k}(b_k-\beta_k)^2\right)\\
&&-c_2\left(\sum_{i}(b_i-\beta_i)\beta_i\right)\left(\sum_{k}(b_k-\beta_k)^4\right)-c_3\left(\sum_{i}(b_i-\beta_i)^5\beta_i\right)\\
&=&O_P\left(1\right),
\end{eqnarray*}
where $c_1,c_2,c_3$ are generic constants.

Similarly we have the following results.
\begin{eqnarray*}
&&\sum_{i\neq j\neq k}
(b_i-\beta_i)(b_j-\beta_j)^2(b_k-\beta_k)\beta_i\beta_k\\
&=&\left(\sum_{i}(b_i-\beta_i)\beta_i\right)\left(\sum_{j}(b_j-\beta_j)^2\right)\left(\sum_{k}(b_k-\beta_k)\beta_k\right)\\
&&-c_1\left(\sum_{i}(b_i-\beta_i)^3\beta_i\right)\left(\sum_{k}(b_k-\beta_k)\beta_k\right)\\
&&-c_2\left(\sum_{i}(b_i-\beta_i)^2\beta_i^2\right)\left(\sum_{j}(b_j-\beta_j)^2\right)-c_3\left(\sum_{i}(b_i-\beta_i)^4\beta_i^2\right)\\
&=&O_P\left(1\right),
\end{eqnarray*}
\begin{eqnarray*}
&&\sum_{i\neq j\neq k}
(b_i-\beta_i)^2(b_j-\beta_j)^2\beta_k^2\leq np_n\left(\sum_{i}(b_i-\beta_i)^2\right)^2=O_P\left(np_n\right),
\end{eqnarray*}
\begin{eqnarray*}
&&\sum_{i\neq j\neq k}
(b_i-\beta_i)(b_j-\beta_j)(b_k-\beta_k)\beta_i\beta_j\beta_k\\
&=&\left(\sum_{i}
(b_i-\beta_i)\beta_i\right)^3-c_1\left(\sum_{i}
(b_i-\beta_i)^2\beta_i^2\right)\left(\sum_{i}
(b_i-\beta_i)\beta_i\right)-c_2\left(\sum_{i}
(b_i-\beta_i)^3\beta_i^3\right)\\
&=&O_P\left(1\right),
\end{eqnarray*}

\begin{eqnarray*}
&&\sum_{i\neq j\neq k}
(b_i-\beta_i)^2(b_j-\beta_j)\beta_j\beta_k^2\\
&=&\left(\sum_{i}
(b_i-\beta_i)^2\right)\left(\sum_{j}
(b_j-\beta_j)\beta_j\right)\sum_k\beta_k^2-c_1\left(\sum_{i}
(b_i-\beta_i)^3\beta_i\right)\sum_k\beta_k^2\\
&&-c_2\left(\sum_{i}
(b_i-\beta_i)^2\beta_k^2\right)\left(\sum_{j}
(b_j-\beta_j)\beta_j\right)-c_3\left(\sum_{i}
(b_i-\beta_i)^2\right)\left(\sum_{j}
(b_j-\beta_j)\beta_j^3\right)\\
&&-c_4\left(\sum_{i}
(b_i-\beta_i)^3\beta_i^3\right)\\
&=&O_P\left(1\right).
\end{eqnarray*}

Then $\sum_{i\neq j\neq k}R_{3,ijk}=o_P(np_n\sqrt{np_n})$.

\qed

\subsubsection{Proof of Lemma \ref{rijksq}}

\noindent
{\bf Proof of Lemma \ref{rijksq}:} We shall prove $\sum_{i\neq j\neq k}R_{t,ijk}^2=o_P((np_n)^3)$ for $t=1,2,3,4,5,6$.

By Lemma \ref{prlem}, we have
\begin{eqnarray}
\sum_{i\neq j\neq k}(b_i-\beta_i)^2(b_j-\beta_j)^2(b_k-\beta_k)^2\beta_i^2\beta_j^2\beta_k^2\leq \left(\sum_{i}(b_i-\beta_i)^2\right)^3=O_P(1),
\end{eqnarray}
and
\begin{eqnarray}
\sum_{i\neq j\neq k}(b_i-\beta_i)^4(b_j-\beta_j)^2\beta_j^2\beta_k^4\leq np_n\left(\sum_{i}(b_i-\beta_i)^4\right)\left(\sum_{i}(b_i-\beta_i)^2\right)=O_P(np_n).
\end{eqnarray}
Then 
\begin{eqnarray}
  \sum_{i\neq j\neq k} R_{6,ijk}^2=o_P\left((np_n)^3\right).
\end{eqnarray}

Since
\begin{eqnarray}
  &&\sum_{i\neq j\neq k}(b_i-\beta_i)^2(b_j-\beta_j)^4(b_k-\beta_k)^2\beta_i^2\beta_k^2\\
  &\leq&p_n^2\left(\sum_{i}(b_i-\beta_i)^2\right)\left(\sum_{j}(b_j-\beta_j)^4\right)\left(\sum_{k}(b_k-\beta_k)^2\right)=O_P(1),
\end{eqnarray}
and
\begin{eqnarray}
\sum_{i\neq j\neq k}(b_i-\beta_i)^4(b_j-\beta_j)^4\beta_k^4\leq np_n^2\left(\sum_{i}(b_i-\beta_i)^4\right)\left(\sum_{j}(b_j-\beta_j)^4\right)=O_P(np_n),
\end{eqnarray}
then 
\begin{eqnarray}
  \sum_{i\neq j\neq k} R_{5,ijk}^2=o_P\left((np_n)^3\right).
\end{eqnarray}

Since
\begin{eqnarray}
  &&\sum_{i\neq j\neq k}(b_i-\beta_i)^2(b_j-\beta_j)^4(b_k-\beta_k)^4\beta_i^2\\
  &\leq& p_n\left(\sum_{i}(b_i-\beta_i)^2\right)\left(\sum_{j}(b_j-\beta_j)^4\right)\left(\sum_{k}(b_k-\beta_k)^4\right)=O_P(1),
\end{eqnarray}
then 
\begin{eqnarray}
  \sum_{i\neq j\neq k} R_{4,ijk}^2=o_P\left((np_n)^3\right).
\end{eqnarray}

Since
\begin{eqnarray}
  &&\sum_{i\neq j\neq k}(b_i-\beta_i)^4(b_j-\beta_j)^4(b_k-\beta_k)^4\\
  &\leq& \left(\sum_{i}(b_i-\beta_i)^4\right)\left(\sum_{j}(b_j-\beta_j)^4\right)\left(\sum_{k}(b_k-\beta_k)^4\right)=O_P(1),
\end{eqnarray}
then 
\begin{eqnarray}
  \sum_{i\neq j\neq k} R_{3,ijk}^2=o_P\left((np_n)^3\right).
\end{eqnarray}

Now we show $\sum_{i\neq j\neq k}R_{2,ijk}^2=o_P\left((np_n)^3\right)$.
By Lemma \ref{prlem}, we have
\begin{eqnarray*}\nonumber
   &&\sum_{i\neq j\neq k} (A_{ij}-\beta_i\beta_j)^2(b_j-\beta_j)^2(b_k-\beta_k)^4(b_i-\beta_i)^2\\ \nonumber
   &\leq&\left(\sum_{i}(b_i-\beta_i)^2\right)\left(\sum_{i}(b_i-\beta_i)^4\right)\left(\sum_{i}(b_i-\beta_i)^2\right)\sum_{i\neq j} (A_{ij}-\beta_i\beta_j)^2\\
   &=&O_P\left(np_n\right),
\end{eqnarray*}
\begin{eqnarray*}\nonumber
   &&\sum_{i\neq j\neq k}(A_{ij}-\beta_i\beta_j)^2(b_j-\beta_j)^2(b_k-\beta_k)^4 \beta_i^2\\
      &\leq&p_n\left(\sum_{j}(b_j-\beta_j)^2\right)\left(\sum_{k}(b_k-\beta_k)^4\right)\sum_{i\neq j} (A_{ij}-\beta_i\beta_j)^2\\
   &=&O_P\left(np_n\right),
\end{eqnarray*}
and
\begin{eqnarray*}\nonumber
   &&\sum_{i\neq j\neq k}
(A_{ij}-\beta_i\beta_j)^2(b_i-\beta_i)^2(b_j-\beta_j)^2(b_k-\beta_k)^2\beta_k^2\\
      &\leq&p_n\left(\sum_{i}(b_i-\beta_i)^2\right)\left(\sum_{j}(b_j-\beta_j)^2\right)\left(\sum_{k}(b_k-\beta_k)^2\right)\sum_{i,j} (A_{ij}-\beta_i\beta_j)^2\\
   &=&O_P\left((np_n)^2\right)
\end{eqnarray*}

The sum of other terms in $\sum_{i\neq j\neq k}R_{2,ijk}$ can be similarly bounded and we get
\begin{eqnarray} \sum_{i\neq j\neq k}R_{2,ijk}^2=o_P\left((np_n)^3\right).
\end{eqnarray}

Now we show $\sum_{i\neq j\neq k}R_{1,ijk}^2=o_P\left((np_n)^3\right)$.
Note that
\begin{eqnarray*} 
  &&\sum_{i\neq j\neq k}(A_{ij}-\beta_i\beta_j)^2(A_{jk}-\beta_j\beta_k)^2(b_k-\beta_k)^2(b_i-\beta_i)^2\\
  &\leq&\left(\sum_{i}(b_i-\beta_i)^2\right)\left[\sum_{i\neq j\neq k}(A_{ij}-\beta_i\beta_j)^2(A_{jk}-\beta_j\beta_k)^2\left(\frac{d_k-\mu_k}{\sqrt{d}}-\frac{(\sqrt{d}-\sqrt{\mu})\mu_k}{\sqrt{d\mu}}\right)^2\right].
\end{eqnarray*}
Recall that $A_{ij}$ are conditionally independent given $Z$ and $\mathbb{E}\big[(\eta_{kt}-\beta_k\beta_t)|Z_k\big]=0$. For $s\neq t$ and $s,t\notin\{i,j,k\}$,  we have
\begin{eqnarray*} 
&&\mathbb{E}\left[(A_{ij}-\beta_i\beta_j)^2(A_{jk}-\beta_j\beta_k)^2(A_{ks}-\beta_k\beta_s)(A_{kt}-\beta_k\beta_t)\right]\\
&=&\mathbb{E}\Big[\mathbb{E}[(A_{ij}-\beta_i\beta_j)^2|Z]\mathbb{E}[(A_{jk}-\beta_j\beta_k)^2|Z]\mathbb{E}[(A_{ks}-\beta_k\beta_s)|Z]\mathbb{E}[(A_{kt}-\beta_k\beta_t)|Z]\Big]\\
&=&\mathbb{E}\Big[\mathbb{E}[(A_{ij}-\beta_i\beta_j)^2|Z]\mathbb{E}[(A_{jk}-\beta_j\beta_k)^2|Z](\eta_{ks}-\beta_k\beta_s)](\eta_{kt}-\beta_k\beta_t)\Big]\\
&=&\mathbb{E}\Bigg[\mathbb{E}\Big[\mathbb{E}[(A_{ij}-\beta_i\beta_j)^2|Z]\mathbb{E}[(A_{jk}-\beta_j\beta_k)^2|Z](\eta_{ks}-\beta_k\beta_s)]|Z_i,Z_j,Z_k,Z_s\Big]\\
&&\times\mathbb{E}\big[(\eta_{kt}-\beta_k\beta_t)|Z_k\big]\Bigg]\\
&=&0.
\end{eqnarray*}
Hence,
\begin{eqnarray*} 
&&\mathbb{E}\left[\sum_{i\neq j\neq k}(A_{ij}-\beta_i\beta_j)^2(A_{jk}-\beta_j\beta_k)^2(d_k-\mu_k)^2\right]\\
&=&\sum_{i\neq j\neq k,s\neq t}\mathbb{E}\left[(A_{ij}-\beta_i\beta_j)^2(A_{jk}-\beta_j\beta_k)^2(A_{ks}-\beta_k\beta_s)(A_{kt}-\beta_k\beta_t)\right]\\
&&+\sum_{i\neq j\neq k,s}\mathbb{E}\left[(A_{ij}-\beta_i\beta_j)^2(A_{jk}-\beta_j\beta_k)^2(A_{ks}-\beta_k\beta_s)^2\right]\\
&=&O(n^4p_n^3).
\end{eqnarray*}
Then
\begin{eqnarray} 
\sum_{i\neq j\neq k}(A_{ij}-\beta_i\beta_j)^2(A_{jk}-\beta_j\beta_k)^2\left(\frac{d_k-\mu_k}{\sqrt{d}}\right)^2=O_P\left((np_n)^2\right).
\end{eqnarray}

It is easy to verify that
\begin{eqnarray} 
\frac{(\sqrt{d}-\sqrt{\mu})^2}{d\mu}\sum_{i\neq j\neq k}(A_{ij}-\beta_i\beta_j)^2(A_{jk}-\beta_j\beta_k)^2\mu_k^2=O_P\left(np_n\right),
\end{eqnarray}
and
\begin{eqnarray} 
  \sum_{i\neq j\neq k}(A_{ij}-\beta_i\beta_j)^2(A_{jk}-\beta_j\beta_k)^2(b_k-\beta_k)^2(b_i-\beta_i)^2=o_P((np_n)^3).
\end{eqnarray}

Then
\begin{eqnarray*} \sum_{i\neq j\neq k}R_{1,ijk}^2=o_P\left((np_n)^3\right).
\end{eqnarray*}

\qed

\subsubsection{Proof of Lemma \ref{lemvar}}

\noindent
{\bf Proof of Lemma \ref{lemvar}:} Straightforward calculation yields
\begin{eqnarray*}
&&\sum_{i\neq j\neq k}(A_{ij}-\beta_i\beta_j)^2(A_{jk}-\beta_j\beta_k)^2(A_{ki}-\beta_k\beta_i)^2\\
&&-\sum_{i\neq j\neq k}\mathbb{E}(A_{ij}-\beta_i\beta_j)^2\mathbb{E}(A_{jk}-\beta_j\beta_k)^2\mathbb{E}(A_{ki}-\beta_k\beta_i)^2\\
&=&\sum_{i\neq j\neq k}\big[(A_{ij}-\beta_i\beta_j)^2-\mathbb{E}(A_{ij}-\beta_i\beta_j)^2\big]\big[(A_{jk}-\beta_j\beta_k)^2-\mathbb{E}(A_{jk}-\beta_j\beta_k)^2\big]\\
&&\times\big[(A_{ki}-\beta_k\beta_i)^2-\mathbb{E}(A_{ki}-\beta_k\beta_i)^2\big]\\
&&+3\sum_{i\neq j\neq k}\big[(A_{ij}-\beta_i\beta_j)^2-\mathbb{E}(A_{ij}-\beta_i\beta_j)^2\big]\big[(A_{jk}-\beta_j\beta_k)^2-\mathbb{E}(A_{jk}-\beta_j\beta_k)^2\big]\mathbb{E}(A_{ki}-\beta_k\beta_i)^2\\
&&+3\sum_{i\neq j\neq k}\big[(A_{ij}-\beta_i\beta_j)^2-\mathbb{E}(A_{ij}-\beta_i\beta_j)^2\big]\mathbb{E}(A_{jk}-\beta_j\beta_k)^2\mathbb{E}(A_{ki}-\beta_k\beta_i)^2\\
&=&O_P\left(\sqrt{(np_n)^3}+\sqrt{n^3p_n^2}+(np_n)^2\right)\\
&=&o_P\left((np_n)^3\right)
\end{eqnarray*}

Hence

\begin{eqnarray*}
&&\frac{1}{(np_n)^3}\sum_{i\neq j\neq k}(A_{ij}-\beta_i\beta_j)^2(A_{jk}-\beta_j\beta_k)^2(A_{ki}-\beta_k\beta_i)^2\\
&=&\frac{1}{(np_n)^3}\sum_{i\neq j\neq k}\mathbb{E}(A_{ij}-\beta_i\beta_j)^2\mathbb{E}(A_{jk}-\beta_j\beta_k)^2\mathbb{E}(A_{ki}-\beta_k\beta_i)^2+o_P(1)\\
&=&\Theta(1)+o_P(1).
\end{eqnarray*}

\qed

\subsubsection{Proof of Lemma \ref{lemmax}}

\noindent
{\bf Proof of Lemma \ref{lemmax}:} By Lemma 2.1 in \cite{A85} and Lemma \ref{lemdmu}, we have
\begin{eqnarray*}
\mathbb{E}\left[\max_{i}\big|d_i-\mu_i\big|^t\right]&\leq& \left(\sum_i\mathbb{E}\big|d_i-\mu_i\big|^{2t}\right)^{\frac{1}{2}}=O\left(\sqrt{n(np_n)^t}\right), \ \ t\leq 5.
\end{eqnarray*}
According to Lemma \ref{plem1}, one has 
\begin{eqnarray*}
\frac{|\sqrt{d}-\sqrt{\mu}|^{m-k}}{\sqrt{d}^{m}\sqrt{\mu}^{m-k}}\max_{i}\big|d_i-\mu_i\big|^k\mu_i^{m-k}=O_P\left(\frac{(np_n)^{m-k}\sqrt{n}(np_n)^{\frac{k}{2}}}{(n^2p_n)^{\frac{2m-k}{2}}}\right)=O_P\left(\frac{\sqrt{n}}{n^{m-\frac{k}{2}}}\right).
\end{eqnarray*}

By the binomial expansion similar to (\ref{r1ijkeq20}), we have
\begin{eqnarray*}
\max_{i}|b_i-\beta_i|^m&=&O_P\left(\frac{1}{n^{\frac{m-1}{2}}}\right).
\end{eqnarray*}
Hence for integers $m_1,m_2,m_3\in\{1,2,3,4,5\}$, we have
\begin{eqnarray*}
&&\max_{i\neq j\neq k}|b_i-\beta_i|^{m_1}|b_j-\beta_j|^{m_2}|b_k-\beta_k|^{m_3}\\
&\leq&\left(\max_{i}|b_i-\beta_i|^{m_1}\right)\left(\max_{j}|b_j-\beta_j|^{m_2}\right)\left(\max_{k}|b_k-\beta_k|^{m_3}\right)=O_P\left(1\right).
\end{eqnarray*}
Then
\begin{eqnarray}\label{maxR}
\max_{i\neq j\neq k}\big|R_{t,ijk}\big|=o_P\left(\sqrt{(np_n)^3}\right),
\end{eqnarray}
for $t=3,4,5,6$.

Next we prove (\ref{maxR}) holds for $t=1,2$. Consider $t=1$ first. Note that
\begin{eqnarray*}
&&\max_{i\neq j\neq k}\big|(A_{ij}-\beta_i\beta_j)(A_{jk}-\beta_j\beta_k)(d_k-\mathbb{E}[d_k])(d_i-\mathbb{E}[d_i])\big|\\   &\leq&\left(\max_{i,j}\big|A_{ij}-\beta_i\beta_j|\right)\left(\max_{j,k}|A_{jk}-\beta_j\beta_k|\right)\left(\max_{k}\big|d_k-\mathbb{E}[d_k]|\right)\left(\max_{i}\big|d_i-\mathbb{E}[d_i]\big|\right).
\end{eqnarray*}

By Lemma 2.1 in \cite{A85}, we have
\begin{eqnarray*}
&&\mathbb{E}\left[\max_{i<j}\big|A_{ij}-\beta_i\beta_j|\right]\leq \left(\sum_{i<j}\mathbb{E}\left[\big|A_{ij}-\beta_i\beta_j|^4\right]\right)^{\frac{1}{4}}=O\left((n^2p_n)^{\frac{1}{4}}\right),
\end{eqnarray*}

\begin{eqnarray*}
&&\mathbb{E}\left[
\max_{k}\big|d_k-\mathbb{E}[d_k]|\right]\leq \left(\sum_{k}\mathbb{E}\left[\big|d_k-\mathbb{E}[d_k]|^4\right]\right)^{\frac{1}{4}}=O\left((n^3p_n^2)^{\frac{1}{4}}\right).
\end{eqnarray*}

Then 
\begin{eqnarray*}
&&\frac{1}{d}\max_{i\neq j\neq k}\big|(A_{ij}-\beta_i\beta_j)(A_{jk}-\beta_j\beta_k)(d_k-\mathbb{E}[d_k])(d_i-\mathbb{E}[d_i])\big|\\
&=&O_P\left(\frac{\sqrt{n^5p_n^3}}{n^2p_n}\right)=o_P\left(\sqrt{(np_n)^3}\right).
\end{eqnarray*}
Similarly, we get
\begin{eqnarray*}
&&\frac{|\sqrt{d}-\sqrt{\mu}|}{d\sqrt{\mu}}\max_{i\neq j\neq k}\big|(A_{ij}-\beta_i\beta_j)(A_{jk}-\beta_j\beta_k)\mu_i(d_k-\mathbb{E}[d_k])\big|\\ &=&O_P\left(\frac{np_n(n^3p_n^2)^{\frac{1}{4} }\sqrt{n^2p_n}}{n^2p_n\sqrt{n^2p_n}}\right)=o_P\left(\sqrt{(np_n)^3}\right).
\end{eqnarray*}

\begin{eqnarray*}
    &&\frac{(\sqrt{d}-\sqrt{\mu})^2}{d\mu}\max_{i\neq j\neq k}\big|(A_{ij}-\beta_i\beta_j)(A_{jk}-\beta_j\beta_k)\mu_k\mu_i\big|\\
&=&O_P\left(\frac{(np_n)^2\sqrt{n^2p_n}}{n^4p_n^2}\right)=o_P\left(\sqrt{(np_n)^3}\right)
\end{eqnarray*}

Hence
\begin{eqnarray*}
\max_{i\neq j\neq k}\big|(A_{ij}-\beta_i\beta_j)(A_{jk}-\beta_j\beta_k)(b_k-\beta_k)(b_i-\beta_i)\big|=o_P\left(\sqrt{(np_n)^3}\right).
\end{eqnarray*}

Note that $\beta_i=O(p_n)$ uniformly for $i$. Hence
\begin{eqnarray*}
&&\max_{i\neq j\neq k}\big|(A_{ij}-\beta_i\beta_j)(A_{jk}-\beta_j\beta_k)(b_k-\beta_k)\beta_i\big|\\
&\leq& \left(\max_{i,j}\big|A_{ij}-\beta_i\beta_j|\right)\left(\max_{j,k}|A_{jk}-\beta_j\beta_k|\right)\left(\max_{k}\big|b_k-\beta_k\big|\right)\sqrt{p_n}\\
&=&O_P\left(np_n\right)=o_P\left(\sqrt{(np_n)^3}\right).
\end{eqnarray*}

Then by (\ref{r1ijkeq1}) we have
\begin{eqnarray}
\max_{i\neq j\neq k}\big|R_{1,ijk}\big|=o_P\left(\sqrt{(np_n)^3}\right).
\end{eqnarray}

Next, we show
\begin{eqnarray}\label{maxR2}
\max_{i\neq j\neq k}\big|R_{2,ijk}\big|=o_P\left(\sqrt{(np_n)^3}\right).
\end{eqnarray}
It is easy to verify that
\begin{eqnarray*}
&&\max_{i\neq j\neq k}\big|(A_{ij}-\beta_i\beta_j)(b_j-\beta_j)(b_k-\beta_k)^2(b_i-\beta_i)\big|\\
&\leq&\left(\max_{i,j}\big|A_{ij}-\beta_i\beta_j|\right)\left(\max_{j}|b_j-\beta_j|\right)\left(\max_{k}(b_k-\beta_k)^2\right)\left(\max_{i}\big|b_i-\beta_i\big|\right)\\
&=&O_P\left(\frac{(n^2p_n)^{\frac{1}{4}}}{\sqrt{n}}\right)=o_P\left(\sqrt{(np_n)^3}\right),
\end{eqnarray*}
\begin{eqnarray*}
&&\max_{i\neq j\neq k}\big|(A_{ij}-\beta_i\beta_j)(b_i-\beta_i)(b_j-\beta_j)(b_k-\beta_k)\beta_k\big|\\
&\leq&\left(\max_{i,j}\big|A_{ij}-\beta_i\beta_j|\right)\left(\max_{j}|b_j-\beta_j|\right)\left(\max_{k}\big|b_k-\beta_k\big|\right)\left(\max_{i}\big|b_i-\beta_i\big|\right)\sqrt{p_n}\\
&=&O_P\left(\sqrt{np_n}\right)=o_P\left(\sqrt{(np_n)^3}\right),
\end{eqnarray*}
and
\begin{eqnarray*}
&&\max_{i\neq j\neq k}\big|(A_{ij}-\beta_i\beta_j)(b_j-\beta_j)(b_k-\beta_k)\beta_i\beta_k\big|\\
&\leq&\left(\max_{i,j}\big|A_{ij}-\beta_i\beta_j|\right)\left(\max_{j}|b_j-\beta_j|\right)\left(\max_{k}\big|b_k-\beta_k\big|\right)p_n\\
&=&O_P\left(\sqrt{np_n}\right)=o_P\left(\sqrt{(np_n)^3}\right).
\end{eqnarray*}
Hence (\ref{maxR2}) holds.

\qed

\subsubsection{Proof of Lemma \ref{lemeta}}

\noindent
{\bf Proof of Lemma \ref{lemeta}:} Straightforward calculation yields
\begin{eqnarray}\nonumber
    &&\sum_{i\neq j\neq k}(A_{ij}-\beta_i\beta_j)(A_{jk}-\beta_j\beta_k)(A_{ki}-\beta_k\beta_i)\\ \nonumber
    &=&\sum_{i\neq j\neq k}(A_{ij}-\eta_{ij}+\eta_{ij}-\beta_i\beta_j)(A_{jk}-\eta_{jk}+\eta_{jk}-\beta_j\beta_k)(A_{ki}-\eta_{ki}+\eta_{ki}-\beta_k\beta_i)\\ \nonumber
    &=&\sum_{i\neq j\neq k}(A_{ij}-\eta_{ij})(A_{jk}-\eta_{jk})(A_{ki}-\eta_{ki})+3\sum_{i\neq j\neq k}(A_{ij}-\eta_{ij})(A_{jk}-\eta_{jk})(\eta_{ki}-\beta_k\beta_i)\\ \nonumber
    &&+3\sum_{i\neq j\neq k}(A_{ij}-\eta_{ij})(\eta_{jk}-\beta_j\beta_k)(\eta_{ki}-\beta_k\beta_i)\\ \label{ssseq20}
    &&+\sum_{i\neq j\neq k}(\eta_{ij}-\beta_i\beta_j)(\eta_{jk}-\beta_j\beta_k)(\eta_{ki}-\beta_k\beta_i).
\end{eqnarray}

It is easy to verify that
\begin{eqnarray*}
\sum_{i\neq j\neq k}(A_{ij}-\eta_{ij})(A_{jk}-\eta_{jk})(A_{ki}-\eta_{ki})=O_P\left(\sqrt{(np_n)^3}\right),
\end{eqnarray*}
and
\begin{eqnarray*}
\sum_{i\neq j\neq k}(A_{ij}-\eta_{ij})(A_{jk}-\eta_{jk})(\eta_{ki}-\beta_k\beta_i)=O_P\left(\sqrt{(np_n)^3}\right).
\end{eqnarray*}

Next we find the  order of the last term of (\ref{ssseq20}).  Direct calculation yields
\begin{eqnarray*}
&&\sum_{i\neq j\neq k}(\eta_{ij}-\beta_i\beta_j)(\eta_{jk}-\beta_j\beta_k)(\eta_{ki}-\beta_k\beta_i)\\
&=&\sum_{i\neq j\neq k}p_n^3\tau^3w_i^2w_j^2w_k^2\big(k_0I[Z_i=Z_j]-1\big)\big(k_0I[Z_j=Z_k]-1\big)\big(k_0I[Z_k=Z_i]-1\big)\\
&=&\sum_{i\neq j\neq k}p_n^3\tau^3w_i^2w_j^2w_k^2\Big((k_0^3-3k_0^2)I[Z_i=Z_j=Z_k]\\
&&+k_0I[Z_i=Z_j]+k_0I[Z_j=Z_k]+k_0I[Z_k=Z_i]-1\Big).
\end{eqnarray*}
Note that 
\begin{eqnarray*}
&&\mathbb{E}\Big[(k_0^3-3k_0^2)I[Z_i=Z_j=Z_k]+k_0I[Z_i=Z_j]+k_0I[Z_j=Z_k]+k_0I[Z_k=Z_i]-1\Big]\\
&=&(k_0^3-3k_0^2)\frac{1}{k_0^2}+3k_0\frac{1}{k_0}-1\\
&=&k_0-1.
\end{eqnarray*}
Then 
\begin{eqnarray*}
\mathbb{E}\Big[\sum_{i\neq j\neq k}(\eta_{ij}-\beta_i\beta_j)(\eta_{jk}-\beta_j\beta_k)(\eta_{ki}-\beta_k\beta_i)\Big]=\Theta(n^3p_n^3\tau^3(k_0-1)).
\end{eqnarray*}

Next we show the last term of (\ref{ssseq20}) is asymptotically equal to its expectation. If $\{i, j, k\}\cap\{i^{\prime},j^{\prime},k^{\prime}\}=\emptyset$, then
\begin{eqnarray*}
&&\mathbb{E}\Bigg[\left(I[Z_i=Z_j=Z_k]-\frac{1}{k_0^2}\right)\left(I[Z_{i^{\prime}}=Z_{j^{\prime}}=Z_{k^{\prime}}]-\frac{1}{k_0^2}\right)\Bigg]=0.
\end{eqnarray*}
Suppose $|\{i, j, k\}\cap\{i^{\prime},j^{\prime},k^{\prime}\}|=1$. Without loss of generality, let $i=i^{\prime}$. Then
\begin{eqnarray*}
&&\mathbb{E}\Bigg[\left(I[Z_i=Z_j=Z_k]-\frac{1}{k_0^2}\right)\left(I[Z_{i^{\prime}}=Z_{j^{\prime}}=Z_{k^{\prime}}]-\frac{1}{k_0^2}\right)\Bigg]\\
&=& \mathbb{E}\Bigg[\left(I[Z_i=Z_j=Z_k=Z_{j^{\prime}}=Z_{k^{\prime}}]-\frac{1}{k_0^4}\right)\Bigg]=0.
\end{eqnarray*}
Suppose $|\{i\neq j\neq k\}\cap\{i^{\prime},j^{\prime},k^{\prime}\}|=2$. Without loss of generality, let $i=i^{\prime}$ and $j=j^{\prime}$. Then
\begin{eqnarray*}
&&\mathbb{E}\Bigg[\left(I[Z_i=Z_j=Z_k]-\frac{1}{k_0^2}\right)\left(I[Z_{i^{\prime}}=Z_{j^{\prime}}=Z_{k^{\prime}}]-\frac{1}{k_0^2}\right)\Bigg]\\
&=& \mathbb{E}\Bigg[\left(I[Z_i=Z_j=Z_k=Z_{k^{\prime}}]-\frac{1}{k_0^4}\right)\Bigg]=\frac{1}{k_0^3}-\frac{1}{k_0^4}.
\end{eqnarray*}
Hence, we get
\begin{eqnarray*}
&&\mathbb{E}\Big[\sum_{i\neq j\neq k}p_n^3\tau^3w_i^2w_j^2w_k^2\Big((k_0^3-3k_0^2)\left(I[Z_i=Z_j=Z_k]-\frac{1}{k_0^2}\right)\Big]^2\\
&=&\sum_{\substack{i\neq j\neq k\\ i^{\prime}\neq j^{\prime}\neq k^{\prime}\\|\{i\neq j\neq k\}\cap\{i^{\prime},j^{\prime},k^{\prime}\}|\geq2
}}p_n^6\tau^6(k_0^3-3k_0^2)^2w_i^2w_j^2w_k^2w_{i^{\prime}}^2w_{j^{\prime}}^2w_{k^{\prime}}^2\\
&&\times \mathbb{E}\left(I[Z_i=Z_j=Z_k=Z_{i^{\prime}}=Z_{j^{\prime}}=Z_{k^{\prime}}]-\frac{1}{k_0^4}\right)\\
&=&O\left(n^4p_n^6\right).
\end{eqnarray*}
Then 
\begin{eqnarray*}
&&\sum_{i\neq j\neq k}p_n^3\tau^3w_i^2w_j^2w_k^2(k_0^3-3k_0^2)I[Z_i=Z_j=Z_k]\\
&=&\sum_{i\neq j\neq k}p_n^3\tau^3w_i^2w_j^2w_k^2(k_0^3-3k_0^2)\mathbb{E}[I[Z_i=Z_j=Z_k]]+O_P(n^2p_n^2).
\end{eqnarray*}

Similarly, we have
\begin{eqnarray*}
\sum_{i\neq j\neq k}p_n^3\tau^3w_i^2w_j^2w_k^2k_0I[Z_i=Z_j]=\sum_{i\neq j\neq k}p_n^3\tau^3w_i^2w_j^2w_k^2k_0\mathbb{E}[I[Z_i=Z_j]]+O_P(n^2p_n^2).
\end{eqnarray*} 

Then we get
\begin{eqnarray*}
\sum_{i\neq j\neq k}(\eta_{ij}-\beta_i\beta_j)(\eta_{jk}-\beta_j\beta_k)(\eta_{ki}-\beta_k\beta_i)&=&(k_0-1)p_n^3\tau^3\sum_{i\neq j\neq k}w_i^2w_j^2w_k^2+O_P(n^2p_n^2)\\
&=&\Theta(n^3p_n^3\tau^3(k_0-1))+O_P(n^2p_n^2).
\end{eqnarray*}

Now we bound the third term of (\ref{ssseq20}). Since
\begin{eqnarray*}
&&\mathbb{E}\Big[\Big(\sum_{i\neq j\neq k}(A_{ij}-\eta_{ij})(\eta_{jk}-\beta_j\beta_k)(\eta_{ki}-\beta_k\beta_i)\Big)^2\Big]\\
&=&\mathbb{E}\Big[\sum_{i\neq j\neq k,k_1}\mathbb{E}[(A_{ij}-\eta_{ij})^2|Z](\eta_{jk}-\beta_j\beta_k)(\eta_{ki}-\beta_k\beta_i)(\eta_{jk_1}-\beta_j\beta_{k_1})(\eta_{k_1i}-\beta_{k_1}\beta_i)\Big]\\
&=&O(n^4p_n^4),
\end{eqnarray*}
then
\begin{eqnarray*}
\sum_{i\neq j\neq k}(A_{ij}-\eta_{ij})(\eta_{jk}-\beta_j\beta_k)(\eta_{ki}-\beta_k\beta_i)=O_P(n^2p_n^2).
\end{eqnarray*}
Then the proof is complete.

\qed

\end{document}